\pdfoutput=1
\documentclass[aps,prb,amsmath,amssymb]{revtex4-1}
\usepackage{graphicx}
\usepackage{epstopdf}
\usepackage{bm}% bold math

\begin{document}

\title[Special Quasi-ordered Structures: role of short-range order in the semiconductor alloy (GaN)$_{1-x}$(ZnO)$_x$]{Special Quasi-ordered Structures: role of short-range order in the semiconductor alloy (GaN)$_{1-x}$(ZnO)$_x$}

\author{Jian Liu}
\email{Jian.Liu@stonybrook.edu}
\author{Maria V. Fern\'{a}ndez-Serra}
\author{Philip B. Allen}
\affiliation{Department of Physics and Astronomy, Stony Brook University, Stony Brook, NY 11794-3800, United States.}
\date{\today}

\begin{abstract}
This paper studies short-range order (SRO) in the semiconductor alloy (GaN)$_{1-x}$(ZnO)$_x$. Monte Carlo simulations performed on a density functional theory (DFT)-based cluster expansion model show that the heterovalent alloys exhibit strong SRO because of the energetic preference for the valence-matched nearest-neighbor Ga-N and Zn-O pairs. To represent the SRO-related structural correlations, we introduce the concept of Special Quasi-ordered Structure (SQoS). Subsequent DFT calculations reveal dramatic influence of SRO on the atomic, electronic and vibrational properties of the (GaN)$_{1-x}$(ZnO)$_x$ alloy. Due to the enhanced statistical presence of the energetically unfavored Zn-N bonds with the strong Zn3$d$-N2$p$ repulsion, the disordered alloys exhibit much larger lattice bowing and band-gap reduction than those of the short-range ordered alloys. Inclusion of lattice vibrations stabilizes the disordered alloy.

\begin{description}
\item[PACS numbers]
\end{description}
\end{abstract}

%Uncomment for PACS numbers title message
\pacs{Valid PACS appear here}

\maketitle

\section{Introduction}
Relative to pure end-member materials, the non-isovalent pseudobinary semiconductor alloy (GaN)$_{1-x}$(ZnO)$_{x}$ shows improved efficiency as a photocatalyst in splitting water into hydrogen and oxygen under visible light illumination\cite{Maeda1}. High efficiency is partly attributed to the band-gap reduction which can be tuned by varying the ZnO content $x$ of the alloy. First-principles calculations show that including short-range order (SRO) affects the band gaps of the heterovalent semiconductor alloys\cite{SiCAlN,ZnSnP}. Experiments on different (GaN)$_{1-x}$(ZnO)$_{x}$ samples also observe large variation in the band gaps, which could be attributed to the different degrees of SRO introduced in growing the samples. For example, the absorption edge shifts monotonically to longer wavelength with increasing $x$ for samples synthesized by nitridation of nanocrystalline ZnGa$_2$O$_4$ and ZnO precursors\cite{Lee}, while a minimum gap at $x\sim0.5$ is found for samples synthesized by mixing of GaN and ZnO powders at high pressure and high temperature\cite{Chen}. Despite the experimental indication of the presence of SRO, a thorough theoretical understanding is still lacking. An even more challenging question is how the vibrational properties depend on SRO, and how they influence the degree of SRO\cite{Ceder,Mookerjee}. The effect of lattice vibrations is important for isovalent semiconductor alloy thermodynamic calculations\cite{SQSphonon}. To our knowledge, to date, there are no published phonon data for (GaN)$_{1-x}$(ZnO)$_{x}$. Therefore we perform detailed first-principles investigations to illuminate the role SRO plays on phonons in the (GaN)$_{1-x}$(ZnO)$_{x}$ alloy.

% It is therefore desirable to explore the SRO effect on the physical properties of (GaN)$_{1-x}$(ZnO)$_{x}$ alloy. Band gap engineering is important in designing photoelectrochemical water splitting materials. 
Previous theoretical studies assume the (GaN)$_{1-x}$(ZnO)$_{x}$ alloy to be completely random\cite{JLL,8Ub,Asymmetric}. The Special Quasirandom Structure (SQS) method\cite{SQS_PRB,SQS_PRL} is often used to construct supercells mimicking random alloys\cite{Bechstedt2002}. However, even for isovalent ternary semiconductor alloys, neglecting SRO introduces non-negligible systematic errors\cite{isovalent,Zunger1994,Zunger1995,Bechstedt2012}. For example, the band gaps of Al$_{0.5}$Ga$_{0.5}$As, Ga$_{0.5}$In$_{0.5}$P and Al$_{0.5}$In$_{0.5}$As alloys can be reduced by as much as 0.1eV through clustering\cite{Zunger1994,Zunger1995}. The electronic properties of wurtzitic Ga$_{1-x}$In$_{x}$N and Al$_{1-x}$In$_{x}$N alloys are also found to be very sensitive to SRO in the cation distribution\cite{Bechstedt2012}. For quaternary alloys the sensitivity to SRO is even greater\cite{SQoSBaroni,SQoSZunger}. The situation is compounded for the (GaN)$_{1-x}$(ZnO)$_{x}$ alloy whose heterovalent nature favors local charge neutrality and therefore valence-matched nearest-neighbor Ga-N and Zn-O pairs. In our previous study on the (GaN)$_{1-x}$(ZnO)$_{x}$ alloy\cite{LL}, referred to hereafter as \uppercase\expandafter{\romannumeral1}, first-principles calculations combined with the cluster expansion method\cite{CE1,CE2,CE3,CEECI} and Monte Carlo simulations predicted a large degree of SRO. In constructing the cluster expansion model, the total energy of a specific configuration is calculated in its relaxed structure. Local relaxations of surprisingly large magnitude are found in our subsequent study\cite{Jian}. The aim of the present study is to construct DFT-affordable supercells whose structural correlations accurately reflect the SRO found by the above approach. The method we use (``special quasi-ordered structure, SQoS'') was used in 1998 by Saitta $et$ $al$\cite{SQoSBaroni} but rarely since then\cite{SQoSZunger}. This method allows us to study with a single DFT calculation, for each $x$, the statistical average atomic, electronic and vibrational properties of the (GaN)$_{1-x}$(ZnO)$_{x}$ alloy. The dependence of structural properties such as bond-length distribution and bond-angle variation upon SRO will be discussed in a separate paper\cite{paper2}.
%Subsequent DFT total-energy and force calculations can then reveal the crucial role of SRO in the atomic, electronic and vibrational properties of the (GaN)$_{1-x}$(ZnO)$_{x}$ alloy.\\

\section{Computational Method}
The (GaN)$_{1-x}$(ZnO)$_{x}$ alloy was modeled in wurtzite structure with interpenetrating cation and anion $hcp$ sublattices. Ga/Zn can only occupy the cation sublattice, while N/O can only occupy the anion sublattice. A detailed description of the cluster expansion model used in this study can be found in \uppercase\expandafter{\romannumeral1}. Monte Carlo simulations are performed using the $ATAT$ package\cite{ATAT1,ATAT2,ATAT3} with a $12\times12\times8$ supercell containing 4608 atoms. For each $(x,T)$, an ensemble of $N$ configurations (labeled by $s=1,2,...,N$) is equilibrated for $1\times10^4$ MC passes followed by a subsequent $1\times10^4$ MC passes sampling. The site occupation is denoted by Ising spin $\sigma_i$ with $\sigma=1$ denoting Ga/N and $\sigma=-1$ denoting Zn/O respectively. For the structural correlations, the notations are adopted from Ref. \onlinecite{SQS_PRB}. The total energy of sample $s$ is expanded in terms of clusters (called ``figures'' and labeled as $\{k,m\}$). The label $k=1,2,...$ is the number of sites of the cluster. The label $m=1,2,...$ enumerates the distinct cluster geometries, as shown in the inset of Fig. 1. The structural correlation function ${\Pi}_{k,m}(l,s)$ describes the occupation of the cluster. The label $(l,s)$ indicates that the cluster is located at location $l$ in sample $s$. For instance, the value of ${\Pi}_{2,1}(l,s)$ is the spin product $\sigma_i\sigma_j$ for the particular nearest-neighbor pair of sites $i$ and $j$ positioned at location $l$ in sample $s$. $\overline{\Pi}_{k,m}(s)$ refers to the average of ${\Pi}_{k,m}(l,s)$ over all locations $l$ in one sample $s$, and $\langle\overline{\Pi}_{k,m}\rangle$ refers to the average of $\overline{\Pi}_{k,m}(s)$ over the samples equilibrated at a certain $(x,T)$.

%A ``figure'' $\{k,m\}$ is defined by the number $k$ of sites ($k=1,2,...$ refer to sites, pairs, etc.) separated by the order $m$ of neighbor distances ($m=1,2,...$ refer to first, second neighbors, etc.) at position $l$ in configuration $s$.
The motivation of the SQS approach\cite{SQS_PRB,SQS_PRL} is to approximate the actual alloy with one representative special structure $S$ whose structural correlation functions $\overline{\Pi}_{k,m}(S)$ best match the corresponding ensemble-averaged $\langle\overline{\Pi}_{k,m}\rangle$ of the realistic alloy\cite{SQS_PRB,SQS_PRL}. The original SQS approach reproduces the average structural correlation functions of the $random$ ($R$) alloy $\overline{\Pi}_{k,m}(S)\sim\langle\overline{\Pi}_{k,m}\rangle_{R}$\cite{SQS_PRB,SQS_PRL}. We extend the SQS approach to the correlation functions of short-range ordered alloys $\langle\overline{\Pi}_{k,m}\rangle_{SRO}$. We first obtain $\langle\overline{\Pi}_{k,m}\rangle_{SRO}$ by performing Monte Carlo simulations on a DFT-based cluster expansion model. Then we generate numerous site occupancies for a certain composition $x$ and look for the representative configuration $S$ for which the set of $\overline{\Pi}_{k,m}(S)$ is closest to $\langle\overline{\Pi}_{k,m}\rangle_{SRO}$ by minimizing $\sum_{k,m}g_{k,m}D_{k,m}\left[\overline{\Pi}_{k,m}(s)-\langle\overline{\Pi}_{k,m}\rangle_{SRO}\right]^2$, where $D_{k,m}$ is the degeneracy (number of equivalent figures) and $g_{k,m}$ is the assigned weighting factor. Enumeration of all possible configurations is not possible since the number grows exponentially with the number of atoms in the supercell. However, increasing the size of the supercell allows better flexibility of matching structural correlation functions. The conflict is eased by the short-range nature of the structural correlations of the (GaN)$_{1-x}$(ZnO)$_{x}$ alloy. The most relevant physical property of the (GaN)$_{1-x}$(ZnO)$_{x}$ alloy is the formation energy, which is dominated by the short-range pair structural correlations\cite{LL}. We thus assign large weighting factors to the nearest-neighbor $\{2,1$-$2\}$ (meaning $\{2,1\}$ and $\{2,2\}$) and next nearest-neighbor $\{2,3$-$6\}$ figures. At each $(x,T)$ we generate $1\times10^5$ 72-atom supercell ($3\times3\times2$) candidate structures among which the best-matching structure $S$ is chosen. We emphasize that the obtained structures are not the optimal SQS. However, the contributions to the energetics from longer-range figures $E(s)-\left<E\right>=\sum_{k,m}D_{k,m}\left[\overline{\Pi}_{k,m}(s)-\langle\overline{\Pi}_{k,m}\rangle_{SRO}\right]\epsilon_{k,m}$ (Eq. (3.1) in Ref. \onlinecite{SQS_PRB}) are reasonably small. To avoid confusion, we name the corresponding special structure for the short-range ordered alloy (equilibrated at the experimental synthesis temperature $T=1,123$K\cite{Maeda1}) as SQoS (Special Quasi-ordered Structure), and for the disordered alloy (equilibrated at an unrealistic high temperature $T=20,000$K) as SQdS (Special Quasi-disordered Structure), in resemblance to the widely used SQS (Special Quasi-random Structure) formalism introduced by Zunger\cite{SQS_PRB,SQS_PRL}. A completely random ``SQS'' at $x=0.5$ is also studied for reference. The constructed special structures are provided in supplementary materials\cite{SM}.
% Only zero-, one- and two-body effective cluster correlations are included, while three-body correlations are found to be unimportant in the original fit. 
% $ATAT$ was used to adjust cluster parameters to fit the formation energy.  The process was iterated, by using the first set of cluster parameters, $via$ MC, to generate new structures, no longer random, but characteristic for temperatures of $\approx$1500K.  New DFT energies and lattice relaxations gave an enlarged and improved data base, which then generated a final set of improved CE parameters

The constructed special structures are fully relaxed with respect to atomic coordinates, volume and shape. Electronic structure calculations are performed using the {\sc Quantum ESPRESSO} package\cite{QE} with the PBEsol functional\cite{PBEsol}. The pseudopotentials are constructed by means of the projector augmented wave (PAW) method\cite{PAW,PSL} with 60 Ry and 240 Ry cutoff energy for plane-wave basis set and charge density respectively. Ga-3$d$ and Zn-3$d$ states are treated explicitly as valence states. The $k$-point mesh is chosen to be equivalent to a $6\times6\times4$ mesh for the 4-atom wurtzite unit cell. To speed the structural relaxations, the input lattice parameters are estimated using Vegard's law\cite{Vegard}. Nowadays fairly large supercells (e.g., over 50 atoms) can be handled at the DFT level. For the non-isovalent semiconductor alloys where large structural relaxations are expected, one can benefit greatly in terms of the computational efficiency from a pre-relaxation prior to the expensive DFT total energy and force calculations. We will address the issue of pre-relaxation in a subsequent study\cite{paper2}. Phonons are calculated using the small displacement method as implemented in the \texttt{PHON} code\cite{PHON}. For each 72-atom primitive cell, a $2\times2\times2$ supercell is used while a small displacement of 0.02 $\AA$ is employed. The force constants are calculated with the {\sc SIESTA} package\cite{SIESTA}. Pseudopotentials for all the atomic species are available from the {\sc SIESTA} homepage\cite{[http://www.icmab.es/siesta/]homepage}, except for Ga a smaller $d$-orbital cutoff radius is used\cite{Ga_pseudo}.

GaN and ZnO have a type-\uppercase\expandafter{\romannumeral2} band alignment\cite{Asymmetric}. The valence band is composed mainly of N-2$p$ states. DFT with LDA or GGA tends to over-delocalize the semicore Zn-$d$ states and consequently over-hybridize the semicore Zn-$d$ states with the N-$p$ states, resulting in an enhancement of the $p$-$d$ repulsion. The band gap is therefore severely underestimated due to the artificially large $p$-$d$ repulsion. In this study we add $U$ corrections to the semicore Ga-$d$ and Zn-$d$ states\cite{Dudarev}. The on-site Coulomb interaction parameter $U\sim 3.1$eV is determined by a first-principles method adopted in Ref. \onlinecite{plusU}. $U$ is approximated as the screened $atomic$ on-site Coulomb interaction $U^{at}/\epsilon_{\infty}$, where $U^{at}$ is the Coulomb energy cost of placing two electrons at the same site ($U^{at}=E_{at}(d^{n+1})+E_{at}(d^{n-1})-2E_{at}(d^{n})$) and $\epsilon_{\infty}$ is the optical (high-frequency) dielectric constant. In this study we take $d^9$ occupancy as the reference point for $d^n$ and evaluate $U^{at}$ from DFT atomic energies. The optical dielectric constant $\epsilon_{\infty}$ is calculated from linear-response theory\cite{linear_response}. A similar approach of screening the exact-exchange by the dielectric constant is shown to improve significantly the performance of the traditional hybrid functionals\cite{alpha}.

\section{Results and Discussions}
\subsection{Structural correlation}
As is predicted in \uppercase\expandafter{\romannumeral1}, the (GaN)$_{1-x}$(ZnO)$_{x}$ alloy is thermodynamically stable over the full range of compositions for $T>870$K. The $x=0.5$ alloy orders at low temperature and undergoes a first-order order-disorder transition at $T\approx 870$K. The ground state is an ordered 50\%-50\% superlattice labeled as (GaN)$_{1}$(ZnO)$_{1}$, where GaN and ZnO double layers stack alternately along the hexagonal $c$-axis ($P6_3mc$). The formation energy for the (GaN)$_{1}$(ZnO)$_{1}$ superlattice is predicted to be small and negative, indicating weak stability against phase separation. An analogous superlattice structure is also predicted for the (SiC)$_{m}$(AlN)$_{n}$ alloy\cite{SiCAlN}.

Upon alloying, the main effect of SRO is to enhance the statistical presence of the valence-matched nearest-neighbor Ga-N and Zn-O pairs. The ensemble-averaged pair correlation functions $\langle\overline{\Pi}_{2,m}\rangle$ at $x=0.5$ (Fig. 1) reveal a large degree of SRO. The nearest-neighbor $\langle\overline{\Pi}_{2,1-2}\rangle$ deviate significantly from the null value of the random alloy, while the next nearest-neighbor $\langle\overline{\Pi}_{2,3-6}\rangle$ are relatively small, comparable with those found in ternary nitride isovalent semiconductor alloys\cite{isovalent}. Longer-range $\langle\overline{\Pi}_{2,7-14}\rangle$ are not important. The long tail of the $\langle\overline{\Pi}_{k,m}\rangle-T$ curve also indicates that SRO persists to high temperature, and therefore complete randomness may not be achievable under common experimental growth conditions. The positive signs of $\langle\overline{\Pi}_{2,1-2}\rangle$ indicate nearest-neighbor preference for the valence-matched Ga-N and Zn-O pairs, while the positive signs of $\langle\overline{\Pi}_{2,3-6}\rangle$ indicate next nearest-neighbor preference for Ga-Ga and Zn-Zn as well as N-N and O-O pairs. The composition dependence of $\langle\overline{\Pi}_{k,m}\rangle$ at $T=1,123$K is shown in Fig. 2. The deviation of $\langle\overline{\Pi}_{k,m}\rangle$ from that of the random alloy increases upon mixing, and yields the largest deviation at $x=0.5$, where neglect of SRO is worst. To compare the degree of SRO included in SQoS, SQdS and SQS, we summarize in Table 1 the corresponding structural correlation functions at $x=0.5$. The 72-atom SQoS, SQdS and SQS accurately reproduce the ensemble-averaged structural correlation functions obtained with a $12\times12\times8$ supercell. These special structures are expected to yield an accurate description of the atomic, electronic and vibrational properties of the (GaN)$_{1-x}$(ZnO)$_{x}$ alloy.
%The effect of SRO on the atomic and electronic structures of the (GaN)$_{1-x}$(ZnO)$_{x}$ alloy derives mainly from the strong energetic preference for valence-matched Ga-N and Zn-O bonds. 
\begin{figure}[htb!]
\centering
\includegraphics[scale=0.3]{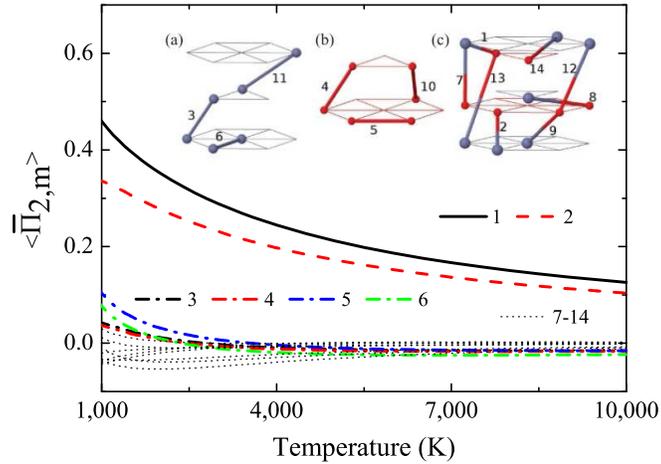}
\caption{Ensemble-averaged pair correlation functions $\langle\overline{\Pi}_{2,m}\rangle$ at $x=0.5$. Definitions of pair figures $\{2,m\}$ can be found in \uppercase\expandafter{\romannumeral1}. $\{2,1$-$2\}$ and $\{2,3$-$6\}$ stand for nearest-neighbor cation-anion pair figures and next nearest-neighbor cation-cation/anion-anion pair figures respectively. Longer-range pair figures $\{2,7$-$14\}$ are shown by dotted lines.}
\end{figure}
\begin{figure}[htb!]
\centering
\includegraphics[scale=0.3]{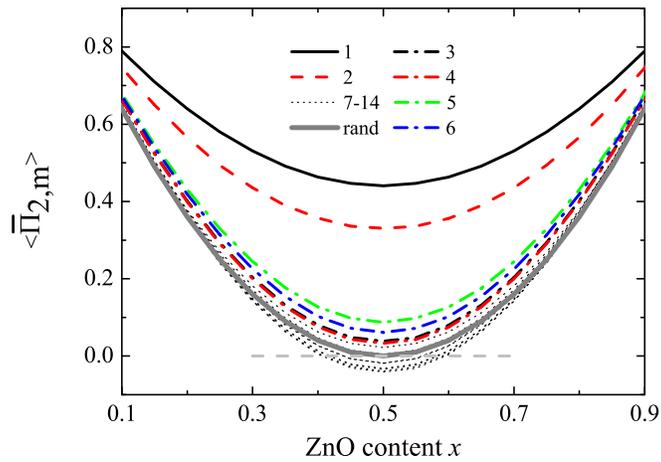}
\caption{Ensemble-averaged pair correlation functions $\langle\overline{\Pi}_{2,m}\rangle$ at $T=1,123$K. The structural correlations for the random alloy $\langle\overline{\Pi}_{k,m}\rangle_{R}=(2x-1)^k$ is shown by the solid grey line for comparison.}
\end{figure}
\begin{table*}
\caption{\label{arttype} 72-atom SQS at $x=0.5$. In spite of the small size of the supercell, the statistical accuracy is good.}
%\begin{indented}
\begin{ruledtabular}
\begin{tabular}{@{}cccccc}
%\br
&$\overline{\Pi}_{2,m}$-SQoS&$\langle\overline{\Pi}_{2,m}\rangle_{1,123K}$&$\overline{\Pi}_{2,m}$-SQdS&$\langle\overline{\Pi}_{2,m}\rangle_{20,000K}$&$\overline{\Pi}_{2,m}$-SQS\\
\colrule
$\{2,1\}$&0.444&0.442&{ }0.074&{ }0.070&0\\
$\{2,2\}$&0.333&0.333&{ }0.000&{ }0.058&0\\
$\{2,3\}$&0.037&0.041&-0.037&-0.011&0\\
$\{2,4\}$&0.037&0.036&{ }0.000&-0.012&0\\
$\{2,5\}$&0.074&0.089&-0.037&-0.012&0\\
$\{2,6\}$&0.074&0.063&{ }0.000&-0.016&0\\
%\br
\end{tabular}
\end{ruledtabular}
%\end{indented}
\end{table*}
\subsection{Atomic, Electronic and Vibrational Properties}
The calculated $U^{at}$, $\epsilon_{\infty}$ and $U$ parameters are listed in Table 2. Compared to the experimental values\cite{property}, the calculated optical dielectric constant is overestimated due to the band-gap underestimation of DFT. However, since the atomic and electronic structures of GaN and ZnO are not very sensitive to the $U$ parameters, the error in the calculated $\epsilon_{\infty}$ (and also the choice of the reference point for $d^n$) does not affect the main conclusions drawn in this study. The calculated lattice constants and band gaps are listed in Table 3. DFT-PBEsol calculations accurately reproduce the lattice constants of GaN and ZnO. The band gap of ZnO is more sensitive to the $U$ correction, due to the strong interaction between the high-lying Zn-3$d$ states and the O-2$p$ states. We then perform DFT+$U$ calculations on the SQoS, SQdS and SQS in order to obtain accurate electronic structure properties. For comparison, total energy and force calculations on configurations randomly selected from the $T=1,123K$ ensembles are also performed within the DFT+$U$ methodology. As shown in Fig. 3, the constructed SQoS accurately represents the ensemble-averaged energetics of the short-range ordered (GaN)$_{1-x}$(ZnO)$_{x}$ alloy. The formation energy of SQoS is significantly lower than that of SQdS. The effect of SRO on the energetics grows upon mixing. Even at $T=20,000$K, the formation energy of SQdS is still considerably lower than that of SQS due to the non-negligible residual SRO.
%The overall improvement compared with LDA of the lattice constants comes at the cost of severely weakening the influence of $U$ corrections on the band gap.
\begin{table}
\caption{\label{arttype} Calculated $U^{at}$, $\epsilon_{\infty}$ and the corresponding $U$ parameters for GaN and ZnO. Experimental values are shown in parenthesis. The PBE version\cite{PBE} of the GGA functional is used instead of PBEsol in obtaining $U^{at}$, due to its better treatment of free atoms.}
%\begin{indented}
\begin{ruledtabular}
\begin{tabular}{@{}cccc}
%\br
&$U^{at}$ (eV)&$\epsilon_{\infty}$&$U$ (eV)\\
\colrule
%\mr
GaN&18.1&5.9 (5.35)&3.1\\
ZnO&16.1&5.2 (3.71)&3.1\\
%\br
\end{tabular}
\end{ruledtabular}
%\end{indented}
\end{table} 
\begin{table}
\caption{\label{arttype} Calculated lattice constants $a$ and $c$ and band gaps $E_g$ for GaN and ZnO.}
%\begin{indented}
\begin{ruledtabular}
\begin{tabular}{@{}ccccccc}
%\br
&\multicolumn{3}{c}{GaN}&\multicolumn{3}{c}{ZnO}\\
&$a$ (\AA)&$c$ (\AA)&$E_{g}$&$a$ (\AA)&$c$ (\AA)&$E_{g}$ (eV)\\
%\mr
\colrule
PBEsol&3.182&5.187&1.88&3.225&5.207&0.71\\
PBEsol+$U$&3.184&5.189&1.89&3.232&5.213&0.92\\
Expt.&3.189&5.185&3.3&3.250&5.204&3.4\\
%\br
\end{tabular}
\end{ruledtabular}
%\end{indented}
\end{table}
\begin{figure}[htb!]
\centering
\includegraphics[scale=0.3]{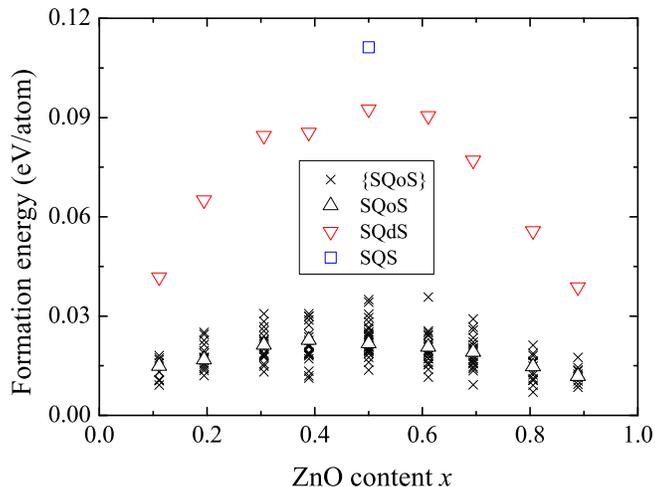}
\caption{DFT-calculated formation energies of SQoS, SQdS and SQS. $\{$SQoS$\}$ is a set of configurations randomly selected from the $T=1,123K$ ensemble.}
\end{figure}

SRO also plays an important role in determining the structural properties. Fig. 4 compares the lattice constant bowing obtained theoretically and experimentally. Once again, the lattice constants of SQoS accurately reproduce the corresponding ensemble-averaged values. With reduced SRO, the disordered alloy shows an expansion as well as a larger bowing compared to the short-range ordered alloy. The experimentally synthesized samples\cite{Lee,Chen} also exhibit moderate bowing, larger than the short-range ordered alloy but smaller than the disordered alloy, indicating the presence of SRO.

Fig. 5 shows the (nearest-neighbor) bond-length distribution of the short-range ordered ($T$=1123K) alloy at $x=0.5$. In the (GaN)$_{1-x}$(ZnO)$_{x}$ alloy, the Ga-N bonds shrink while the Zn-O bonds expand. This unusual bond-length distribution is determined by the non-isovalent nature of the alloy. A follow-up study\cite{paper2} will discuss the prediction and explanation of the bond-length distribution based on the concept of bond valence\cite{BVM}. For the (GaN)$_{1-x}$(ZnO)$_{x}$ alloy, the Zn-N bond-length distribution has crucial importance since it is related to the band-gap reduction through the Zn3$d$-N2$p$ repulsion. In Fig. 6 we show its dependence on the ZnO content. We find that the Zn-N bond-length distribution of the short-range ordered alloy shifts to shorter bonds as the ZnO content increases. Shorter Zn-N bond lengths result in stronger Zn$3d$-N$2p$ repulsion and therefore significantly push up the top of the valence band.

The bond-angle variation is also unusual, namely N-Ga-N and Ga-N-Ga angles expand while O-Zn-O and Zn-O-Zn angles shrink relative to the ideal tetrahedral angle 109.5$^{\circ}$. Fig. 7 shows the variation of bond angles. For example, the Ga centered bond angle shrinks with increased presence of ligand O atoms. This tendency can also be explained using the concept of bond valence. For Fig. 5-7, see Ref. \onlinecite{paper2} for a statistically reliable prediction based on the bond valence method.
\begin{figure}[htb!]
\centering
\includegraphics[scale=0.3]{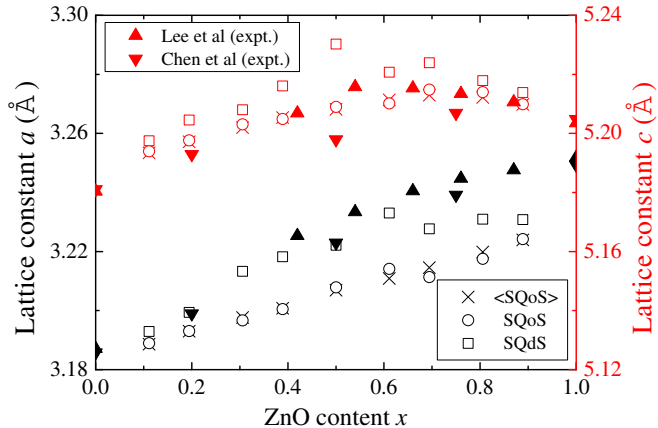}
\caption{DFT-calculated lattice constants of SQoS and SQdS.}
\end{figure}
\begin{figure}[htb!]
\centering
\includegraphics[scale=0.3]{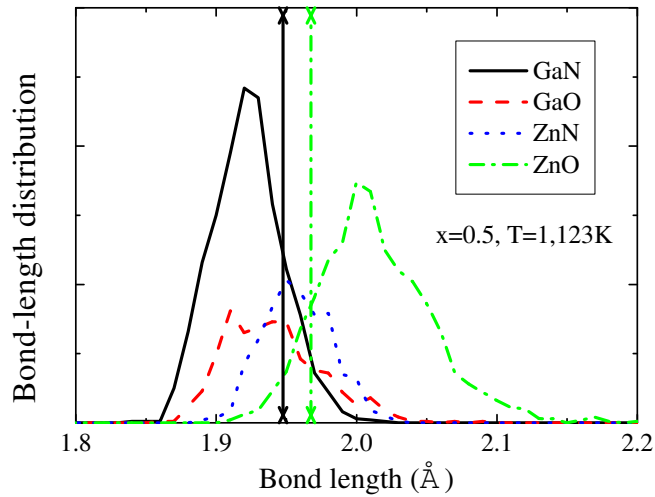}
\caption{DFT-calculated bond-length distribution at $T=1123$K and $x=0.5$. Thirty 72-atom structures are selected from the corresponding thermodynamic ensemble. The bin interval is set to 0.01\AA. The vertical lines mark the bond lengths of the corresponding compounds.}
\end{figure}
\begin{figure}[htb!]
\centering
\includegraphics[scale=0.3]{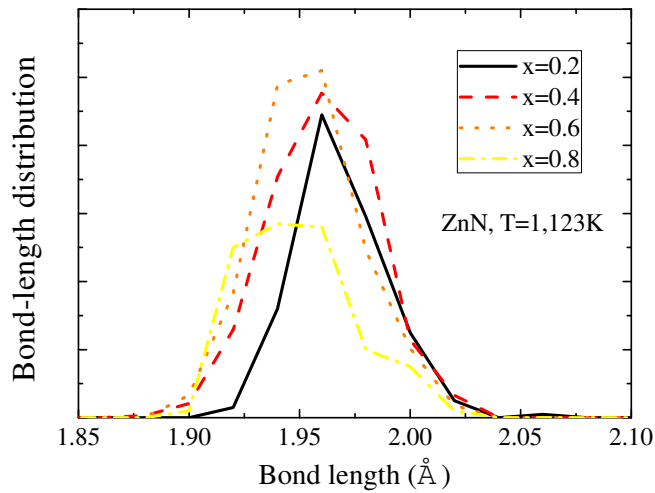}
\caption{DFT-calculated Zn-N bond-length distribution at $T=1123$K.}
\end{figure}
\begin{figure}[htb!]
\centering
\includegraphics[scale=0.3]{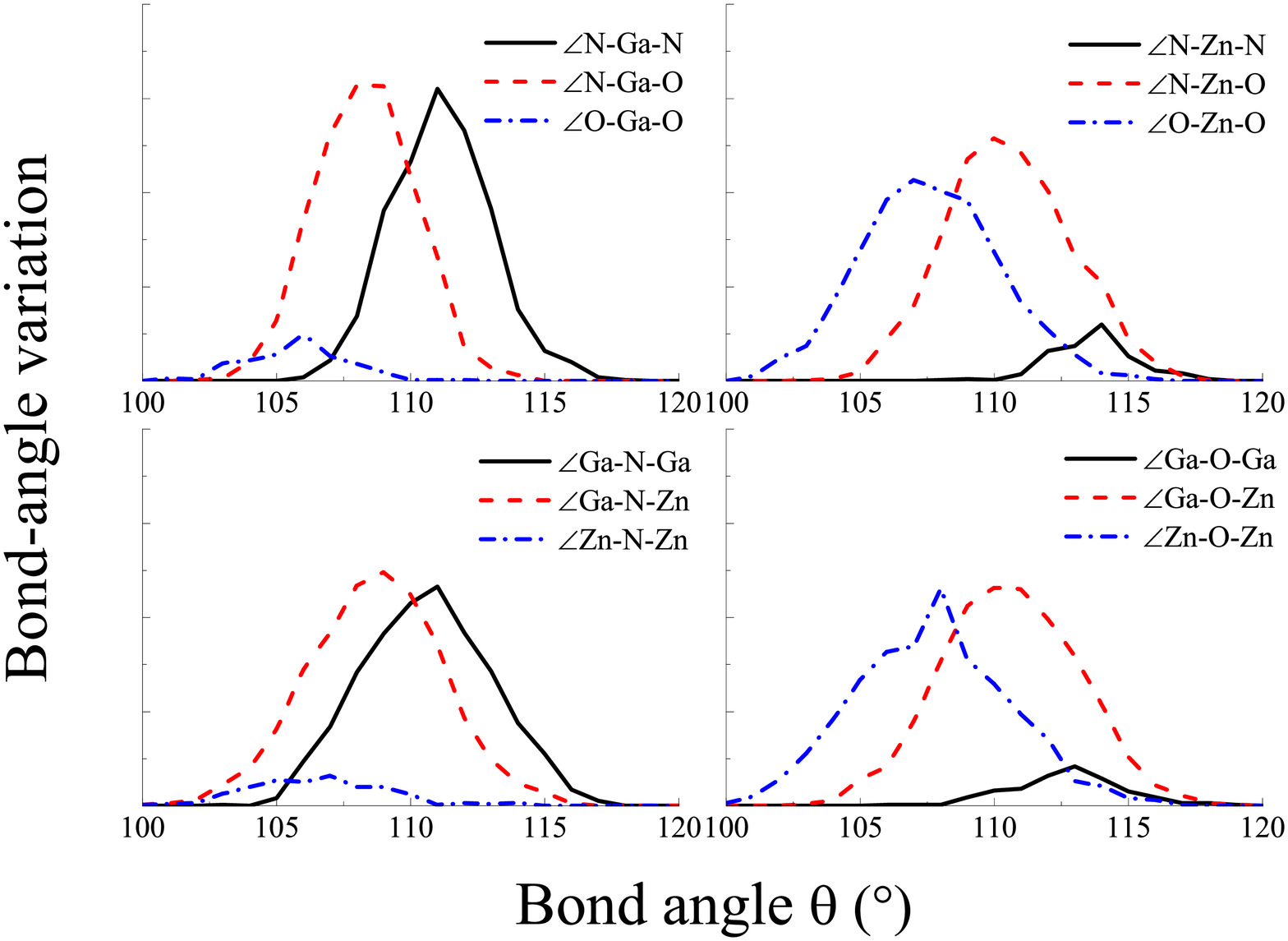}
\caption{DFT-calculated bond-angle variation at $(T=1123$K$,x=0.5$).}
\end{figure}

The atomic and electronic structures of the short-range ordered alloys deviate significantly from those of the disordered alloys. Theoretical atomistic modeling requires explicit inclusion of SRO. Fig. 8 shows the calculated band gaps of SQoS and SQdS. Due to the enhanced statistical presence of the Zn-N bonds, the band gap of the disordered alloy is further reduced relative to that of the short-range ordered alloy. The band-gap reduction is asymmetric. For the disordered alloys the band gap bowing is somewhat parabolic, while for the short-range ordered alloys the band gap reduces almost linearly with increasing ZnO in the GaN host. The linear band-gap reduction is maintained even for the unrelaxed short-range ordered alloys, indicating the dominating role of configurational SRO. In Fig. 8 we also show the linear redshift of the absorption onset with increased ZnO content observed in samples synthesized by nitridation of nanocrystalline ZnGa$_2$O$_4$ and ZnO precursors\cite{Lee}. The linearity is a clear indication of the presence of SRO. We also notice that the high-temperature and high-pressure synthesized samples exhibit the minimum gap at $x=0.5$\cite{Chen}, which is somewhat consistent with the parabolic band gap bowing of the random alloy. The parabolic bowing is attributed to the promoted kinetics of mixing at high-temperature and high-pressure. The contrast in the band gap bowing is a clear indication of the importance of SRO. Since the SRO introduced in the sample is related to the synthesis techniques and the growth conditions, one might therefore consider the opportunity of engineering the band gap $E_g(x,T,\Pi)$ via SRO.
\begin{figure}[htb!]
\centering
\includegraphics[scale=0.3]{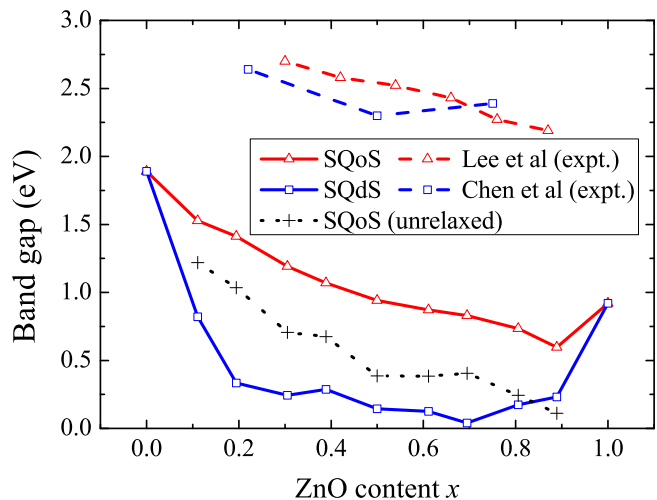}
\caption{DFT-calculated band gaps of SQoS and SQdS. Experimental measurements (Lee et al in Ref. \onlinecite{Lee} and Chen et al in Ref. \onlinecite{Chen}) are also shown for comparison.}
\end{figure}

Fig. 9 compares the projected density of states (PDOS) of SQoS and SQdS at $x=0.5$. The main contribution to the bottom of the valence band comes from O-2$s$ states, which are taken as the reference level because they are less sensitive to the local chemical environment. The top of the valence band is mainly composed of N-2$p$ states. For the disordered alloy the increased statistical presence of the energetically unfavored Zn-N pairs pushes the band edge upward, resulting in further reduction of the band gap. The N-2$p$ states depend strongly on the local chemical environment. Fig. 10 shows the PDOS of N-2$p$ states with the N atoms surrounded by different numbers of Zn atoms. The N-2$p$ states shift upward (dashed lines in Fig. 10) with increased presence of Zn neighbors.
\begin{figure}[htb!]
\centering
\includegraphics[scale=0.3]{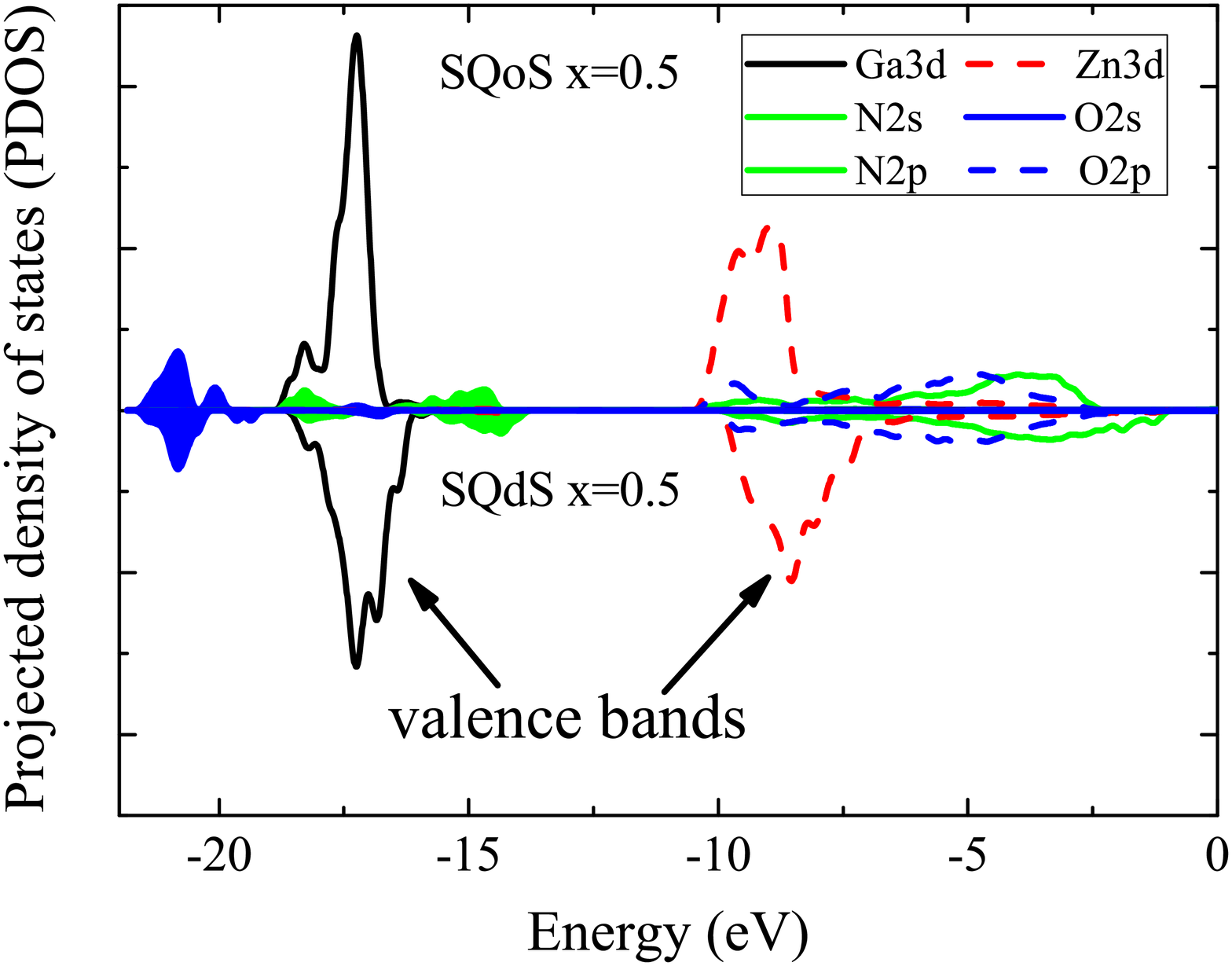}
\caption{Projected density of states (PDOS) of the valence band. The cation-$s$ states in the conduction band are not shown. The deep-lying anion-$s$ states are shown by the shaded area.}
\end{figure}
\begin{figure}[htb!]
\centering
\includegraphics[scale=0.3]{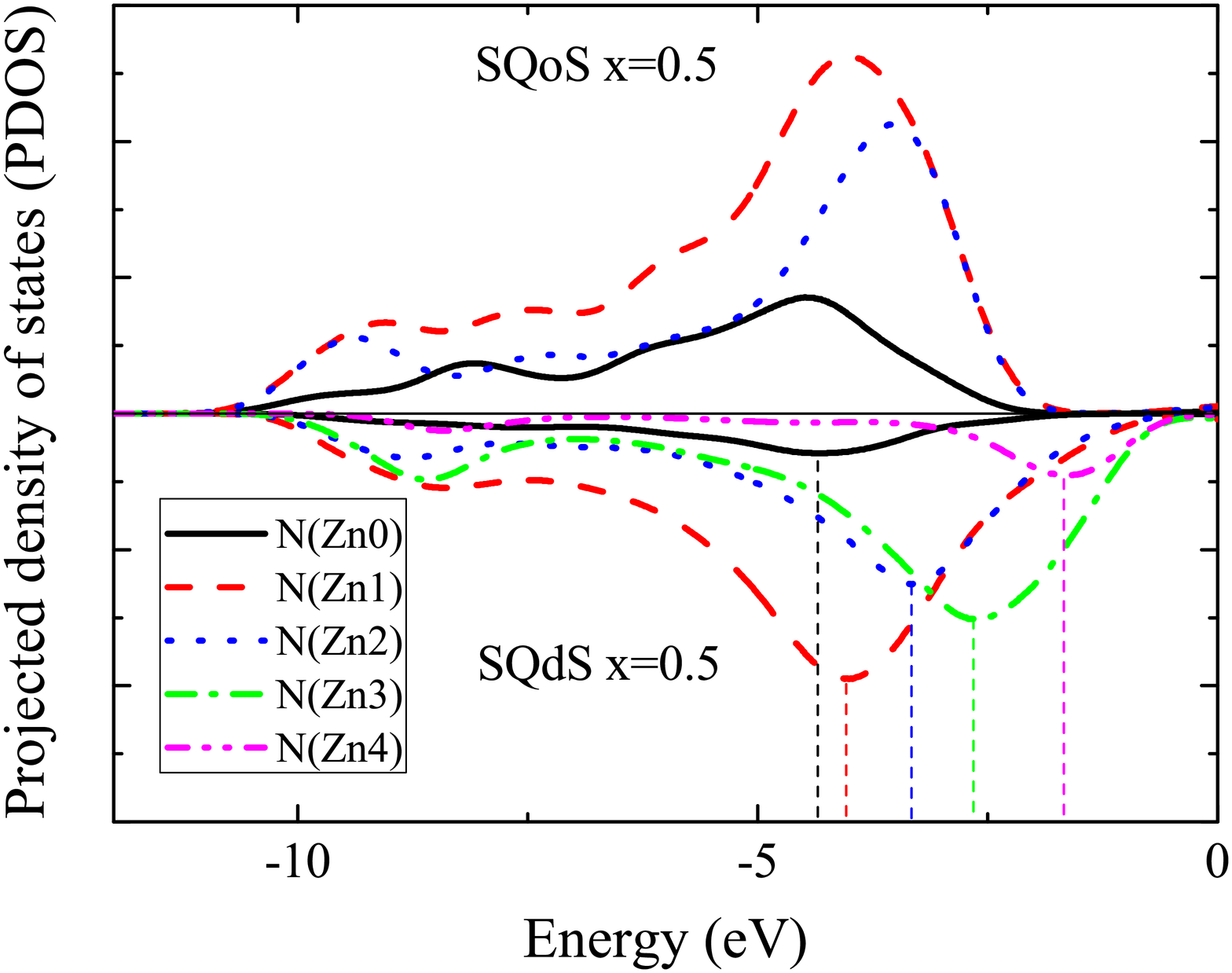}
\caption{PDOS of N-2$p$ with N atoms being surrounded by 0, 1, 2, 3 or 4 Zn neighbors. PDOS is not normalized so that the area under each curve represents the corresponding statistical presence.}
\end{figure}

The effect of lattice vibrations is calculated using the harmonic approximation. The phonon DOS for the SQoS alloys along with those of compound GaN and ZnO are shown in Fig. 11. Three mechanisms have been suggested to explain the origin of vibrational entropy differences in alloys\cite{RMP}: the $bond$ $proportion$ effect, the $volume$ effect and the $size$ $mismatch$ effect. Upon disordering, the $bond$ $proportion$ effect is typically associated with a broadening of the phonon DOS due to the statistical presence of bonds with different stiffness. On the other hand, the $volume$ effect is usually characterized by an overall shift of the phonon DOS due to the change in the frequency of all phonon modes. For the (GaN)$_{1-x}$(ZnO)$_{x}$ alloy, as the alloy expands with increasing $x$, the phonon DOS shifts to lower frequencies as the chemical bonds are in general softened. The $volume$ effect is magnified by the fact that the ``ionic'' Zn-O bond is softer than the ``covalent'' Ga-N bond. The low-frequency phonon DOS is well represented by the composition weighted average $(1-x)g_{\rm{GaN}}(\omega)$+$xg_{\rm{ZnO}}(\omega)$. A significant part of the phonon DOS difference (and therefore the vibrational entropy difference) comes from the high-frequency phonons. The effect of SRO is shown for the $x=0.5$ case. The high-frequency phonon DOS of the SQdS exhibits a much broader spectrum than that of the SQoS. Consequently the phonon mixing entropy of the SQdS is three times larger than that of the SQoS, as is shown in Fig. 12. While the $x$-dependence of the configurational mixing entropy is symmetric\cite{Jian}, the $x$-dependence of the phonon mixing entropy is highly asymmetric, indicating that the inclusion of the vibrational free energy into the alloy thermodynamics could alter the shape of the phase diagram.
\begin{figure}[htb!]
\centering
\includegraphics[scale=0.3]{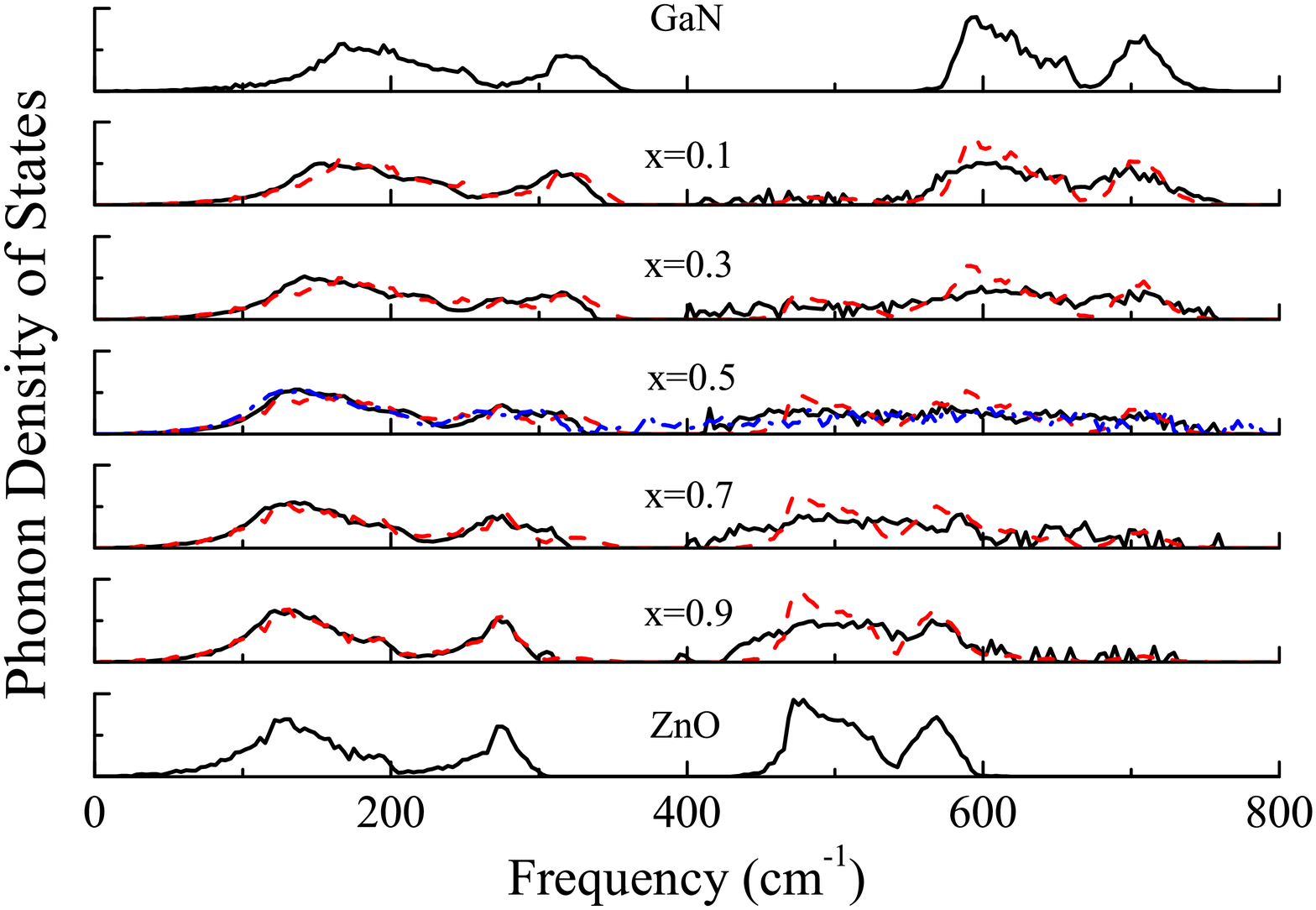}
\caption{Phonon DOS for the SQoS (solid black) ($x=0.1,0.3,0.5,0.7$ and $0.9$) and SQdS (dash-dot blue) ($x=0.5$) alloys. The dash red lines represent the corresponding average of phonon DOS: $(1-x)g_{\rm{GaN}}(\omega)$+$xg_{\rm{ZnO}}(\omega)$.}
\end{figure}
\begin{figure}[htb!]
\centering
\includegraphics[scale=0.3]{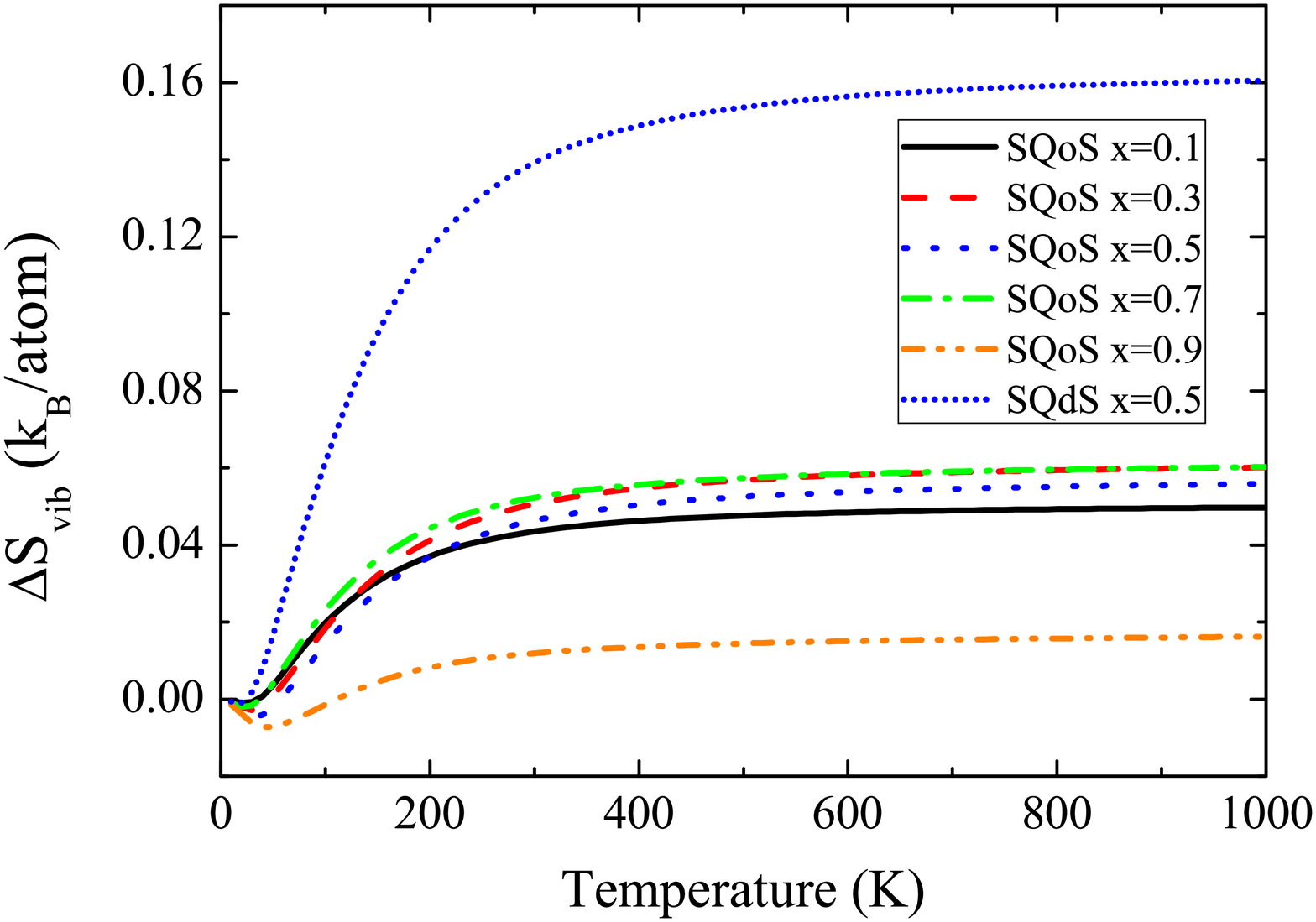}
\caption{$x$-dependence of the phonon mixing entropy $\Delta S_{vib}=S_{vib}(x)-[(1-x)S_{vib}^{\rm{GaN}}+xS_{vib}^{\rm{ZnO}}]$.}
\end{figure}
\section{Conclusions}
The importance of SRO in atomistic modeling schemes such as the SQS approach is often overlooked. For binary metal alloys or isovalent semiconductor alloys, SRO is usually less important. However, for the non-isovalent semiconductor alloys, the valence-matching driving force induces significant SRO. The SQS approach provides a way of approximating the actual alloy with a DFT-affordable supercell. In order to properly compute the non-isovalent alloy, one needs prior knowledges of SRO. In this study the correlated site occupations are provided by Monte Carlo simulations on a DFT-based cluster expansion model. Exhaustive enumeration of all site occupations is avoided due to the SRO in the (GaN)$_{1-x}$(ZnO)$_{x}$ alloy. We seek to match only the cation-anion nearest-neighbor and the cation-cation/anion-anion next nearest-neighbor correlations. The longer-range correlations are optimized to a lesser extent. Since the short-range nature is generic in the non-isovalency, the construction of SQoS proposed in the present study should also be applicable to other non-isovalent semiconductor alloys. If longer-range correlations come into play, one might apply for example the evolutionary algorithm in order to efficiently search for the optimal SQoS.
%A large supercell is necessary in order to include distinguishable degrees of SRO.

The present study reveals the presence of strong SRO in the (GaN)$_{1-x}$(ZnO)$_{x}$ alloy. We construct reliable SQoS and SQdS whose structural correlations reproduce those of the short-range ordered alloys and the disordered alloys respectively. Atomic, electronic and vibrational properties of the short-range ordered alloys deviate significantly from those of the disordered alloys. The short-range ordered alloys experience smaller lattice bowing than the disordered alloys. We offer a tentative explanation in terms of SRO for the discrepancy of the band gaps found in samples synthesized by different methods. SRO inhibits the nearest-neighbor Zn-N pairs, which affects the strength of the Zn3$d$-N2$p$ repulsion and consequently the top of the valence band. The dependence of the N-2$p$ states on local chemical environment demonstrates the vital role of SRO in accurately describing the (GaN)$_{1-x}$(ZnO)$_{x}$ alloy. Phonon DOS is sensitive to the presence of SRO. Disordered alloys have much larger vibrational entropy of mixing than short-range ordered alloys.

\begin{acknowledgments}
This research used computational resources at the Center for Functional Nanomaterials, Brookhaven National Laboratory, which is supported by the US Department of Energy under Contract No. DE-AC02-98CH10886. Work at Stony Brook was supported by US DOE Grant No. DE-FG02-08ER46550 (PBA) and DE-FG02-09ER16052 (MFS). Jian Liu is also sponsored by the China Scholarship Council (CSC).
\end{acknowledgments}

%\bibitem{Chen} Chen H, Wang L, Bai J,Hanson J C, Warren J B, Muckerman J T, Fujita E and Rodriguez J A 2010 \textit{J. Phys. Chem. C} \textbf{114} 1809
%\bibitem{Maeda1} Maeda K, Takata T, Hara M, Saito N, Inoue Y, Kobayashi H and Domen K 2005 \textit{J. Am. Chem. Soc.} \textbf{127} 8286
%\bibitem{Lee} Lee K, Tienes B M, Wilker M B, Schnitzenbaumer K J and Dukovic G 2012 \textit{Nano Lett.} \textbf{12} 3268
%\bibitem{Maeda2} Maeda K, Teramura K, Lu D L, Takata T, Saito N, Inoue Y and Domen K 2006 \textit{Nature} \textbf{440} 295
%\section*{References}

\nocite{*}

\bibliography{SQoS}

%merlin.mbs apsrev4-1.bst 2010-07-25 4.21a (PWD, AO, DPC) hacked
%Control: key (0)
%Control: author (8) initials jnrlst
%Control: editor formatted (1) identically to author
%Control: production of article title (-1) disabled
%Control: page (0) single
%Control: year (1) truncated
%Control: production of eprint (0) enabled
\begin{thebibliography}{48}%
\makeatletter
\providecommand \@ifxundefined [1]{%
 \@ifx{#1\undefined}
}%
\providecommand \@ifnum [1]{%
 \ifnum #1\expandafter \@firstoftwo
 \else \expandafter \@secondoftwo
 \fi
}%
\providecommand \@ifx [1]{%
 \ifx #1\expandafter \@firstoftwo
 \else \expandafter \@secondoftwo
 \fi
}%
\providecommand \natexlab [1]{#1}%
\providecommand \enquote  [1]{``#1''}%
\providecommand \bibnamefont  [1]{#1}%
\providecommand \bibfnamefont [1]{#1}%
\providecommand \citenamefont [1]{#1}%
\providecommand \href@noop [0]{\@secondoftwo}%
\providecommand \href [0]{\begingroup \@sanitize@url \@href}%
\providecommand \@href[1]{\@@startlink{#1}\@@href}%
\providecommand \@@href[1]{\endgroup#1\@@endlink}%
\providecommand \@sanitize@url [0]{\catcode `\\12\catcode `\$12\catcode
  `\&12\catcode `\#12\catcode `\^12\catcode `\_12\catcode `\%12\relax}%
\providecommand \@@startlink[1]{}%
\providecommand \@@endlink[0]{}%
\providecommand \url  [0]{\begingroup\@sanitize@url \@url }%
\providecommand \@url [1]{\endgroup\@href {#1}{\urlprefix }}%
\providecommand \urlprefix  [0]{URL }%
\providecommand \Eprint [0]{\href }%
\providecommand \doibase [0]{http://dx.doi.org/}%
\providecommand \selectlanguage [0]{\@gobble}%
\providecommand \bibinfo  [0]{\@secondoftwo}%
\providecommand \bibfield  [0]{\@secondoftwo}%
\providecommand \translation [1]{[#1]}%
\providecommand \BibitemOpen [0]{}%
\providecommand \bibitemStop [0]{}%
\providecommand \bibitemNoStop [0]{.\EOS\space}%
\providecommand \EOS [0]{\spacefactor3000\relax}%
\providecommand \BibitemShut  [1]{\csname bibitem#1\endcsname}%
\let\auto@bib@innerbib\@empty
%</preamble>
\bibitem [{\citenamefont {Maeda}\ \emph {et~al.}(2005)\citenamefont {Maeda},
  \citenamefont {Takata}, \citenamefont {Hara}, \citenamefont {Saito},
  \citenamefont {Inoue}, \citenamefont {Kobayashi},\ and\ \citenamefont
  {Domen}}]{Maeda1}%
  \BibitemOpen
  \bibfield  {author} {\bibinfo {author} {\bibfnamefont {K.}~\bibnamefont
  {Maeda}}, \bibinfo {author} {\bibfnamefont {T.}~\bibnamefont {Takata}},
  \bibinfo {author} {\bibfnamefont {M.}~\bibnamefont {Hara}}, \bibinfo {author}
  {\bibfnamefont {N.}~\bibnamefont {Saito}}, \bibinfo {author} {\bibfnamefont
  {Y.}~\bibnamefont {Inoue}}, \bibinfo {author} {\bibfnamefont
  {H.}~\bibnamefont {Kobayashi}}, \ and\ \bibinfo {author} {\bibfnamefont
  {K.}~\bibnamefont {Domen}},\ }\href {\doibase 10.1021/ja0518777} {\bibfield
  {journal} {\bibinfo  {journal} {J. Am. Chem. Soc.}\ }\textbf {\bibinfo
  {volume} {127}},\ \bibinfo {pages} {8286} (\bibinfo {year}
  {2005})}\BibitemShut {NoStop}%
\bibitem [{\citenamefont {Burton}\ \emph {et~al.}(2011)\citenamefont {Burton},
  \citenamefont {Demers},\ and\ \citenamefont {van~de Walle}}]{SiCAlN}%
  \BibitemOpen
  \bibfield  {author} {\bibinfo {author} {\bibfnamefont {B.~P.}\ \bibnamefont
  {Burton}}, \bibinfo {author} {\bibfnamefont {S.}~\bibnamefont {Demers}}, \
  and\ \bibinfo {author} {\bibfnamefont {A.}~\bibnamefont {van~de Walle}},\
  }\href {\doibase http://dx.doi.org/10.1063/1.3602149} {\bibfield  {journal}
  {\bibinfo  {journal} {J. Appl. Phys.}\ }\textbf {\bibinfo {volume} {110}},\
  \bibinfo {eid} {023507} (\bibinfo {year} {2011})}\BibitemShut {NoStop}%
\bibitem [{\citenamefont {Ma}\ \emph {et~al.}(2014)\citenamefont {Ma},
  \citenamefont {Deng}, \citenamefont {Luo},\ and\ \citenamefont
  {Wei}}]{ZnSnP}%
  \BibitemOpen
  \bibfield  {author} {\bibinfo {author} {\bibfnamefont {J.}~\bibnamefont
  {Ma}}, \bibinfo {author} {\bibfnamefont {H.-X.}\ \bibnamefont {Deng}},
  \bibinfo {author} {\bibfnamefont {J.-W.}\ \bibnamefont {Luo}}, \ and\
  \bibinfo {author} {\bibfnamefont {S.-H.}\ \bibnamefont {Wei}},\ }\href
  {\doibase 10.1103/PhysRevB.90.115201} {\bibfield  {journal} {\bibinfo
  {journal} {Phys. Rev. B}\ }\textbf {\bibinfo {volume} {90}},\ \bibinfo
  {pages} {115201} (\bibinfo {year} {2014})}\BibitemShut {NoStop}%
\bibitem [{\citenamefont {Lee}\ \emph {et~al.}(2012)\citenamefont {Lee},
  \citenamefont {Tienes}, \citenamefont {Wilker}, \citenamefont
  {Schnitzenbaumer},\ and\ \citenamefont {Dukovic}}]{Lee}%
  \BibitemOpen
  \bibfield  {author} {\bibinfo {author} {\bibfnamefont {K.}~\bibnamefont
  {Lee}}, \bibinfo {author} {\bibfnamefont {B.~M.}\ \bibnamefont {Tienes}},
  \bibinfo {author} {\bibfnamefont {M.~B.}\ \bibnamefont {Wilker}}, \bibinfo
  {author} {\bibfnamefont {K.~J.}\ \bibnamefont {Schnitzenbaumer}}, \ and\
  \bibinfo {author} {\bibfnamefont {G.}~\bibnamefont {Dukovic}},\ }\href
  {\doibase 10.1021/nl301338z} {\bibfield  {journal} {\bibinfo  {journal} {Nano
  Lett.}\ }\textbf {\bibinfo {volume} {12}},\ \bibinfo {pages} {3268} (\bibinfo
  {year} {2012})}\BibitemShut {NoStop}%
\bibitem [{\citenamefont {Chen}\ \emph {et~al.}(2010)\citenamefont {Chen},
  \citenamefont {Wang}, \citenamefont {Bai}, \citenamefont {Hanson},
  \citenamefont {Warren}, \citenamefont {Muckerman}, \citenamefont {Fujita},\
  and\ \citenamefont {Rodriguez}}]{Chen}%
  \BibitemOpen
  \bibfield  {author} {\bibinfo {author} {\bibfnamefont {H.}~\bibnamefont
  {Chen}}, \bibinfo {author} {\bibfnamefont {L.}~\bibnamefont {Wang}}, \bibinfo
  {author} {\bibfnamefont {J.}~\bibnamefont {Bai}}, \bibinfo {author}
  {\bibfnamefont {J.~C.}\ \bibnamefont {Hanson}}, \bibinfo {author}
  {\bibfnamefont {J.~B.}\ \bibnamefont {Warren}}, \bibinfo {author}
  {\bibfnamefont {J.~T.}\ \bibnamefont {Muckerman}}, \bibinfo {author}
  {\bibfnamefont {E.}~\bibnamefont {Fujita}}, \ and\ \bibinfo {author}
  {\bibfnamefont {J.~A.}\ \bibnamefont {Rodriguez}},\ }\href {\doibase
  10.1021/jp909649n} {\bibfield  {journal} {\bibinfo  {journal} {J. Phys. Chem.
  C}\ }\textbf {\bibinfo {volume} {114}},\ \bibinfo {pages} {1809} (\bibinfo
  {year} {2010})}\BibitemShut {NoStop}%
\bibitem [{\citenamefont {van~de Walle}\ and\ \citenamefont
  {Ceder}(2000)}]{Ceder}%
  \BibitemOpen
  \bibfield  {author} {\bibinfo {author} {\bibfnamefont {A.}~\bibnamefont
  {van~de Walle}}\ and\ \bibinfo {author} {\bibfnamefont {G.}~\bibnamefont
  {Ceder}},\ }\href {\doibase 10.1103/PhysRevB.61.5972} {\bibfield  {journal}
  {\bibinfo  {journal} {Phys. Rev. B}\ }\textbf {\bibinfo {volume} {61}},\
  \bibinfo {pages} {5972} (\bibinfo {year} {2000})}\BibitemShut {NoStop}%
\bibitem [{\citenamefont {Alam}\ \emph {et~al.}(2011)\citenamefont {Alam},
  \citenamefont {Chouhan},\ and\ \citenamefont {Mookerjee}}]{Mookerjee}%
  \BibitemOpen
  \bibfield  {author} {\bibinfo {author} {\bibfnamefont {A.}~\bibnamefont
  {Alam}}, \bibinfo {author} {\bibfnamefont {R.~K.}\ \bibnamefont {Chouhan}}, \
  and\ \bibinfo {author} {\bibfnamefont {A.}~\bibnamefont {Mookerjee}},\ }\href
  {\doibase 10.1103/PhysRevB.83.054201} {\bibfield  {journal} {\bibinfo
  {journal} {Phys. Rev. B}\ }\textbf {\bibinfo {volume} {83}},\ \bibinfo
  {pages} {054201} (\bibinfo {year} {2011})}\BibitemShut {NoStop}%
\bibitem [{\citenamefont {Gan}\ \emph {et~al.}(2006)\citenamefont {Gan},
  \citenamefont {Feng},\ and\ \citenamefont {Srolovitz}}]{SQSphonon}%
  \BibitemOpen
  \bibfield  {author} {\bibinfo {author} {\bibfnamefont {C.~K.}\ \bibnamefont
  {Gan}}, \bibinfo {author} {\bibfnamefont {Y.~P.}\ \bibnamefont {Feng}}, \
  and\ \bibinfo {author} {\bibfnamefont {D.~J.}\ \bibnamefont {Srolovitz}},\
  }\href {\doibase 10.1103/PhysRevB.73.235214} {\bibfield  {journal} {\bibinfo
  {journal} {Phys. Rev. B}\ }\textbf {\bibinfo {volume} {73}},\ \bibinfo
  {pages} {235214} (\bibinfo {year} {2006})}\BibitemShut {NoStop}%
\bibitem [{\citenamefont {Jensen}\ \emph {et~al.}(2008)\citenamefont {Jensen},
  \citenamefont {Muckerman},\ and\ \citenamefont {Newton}}]{JLL}%
  \BibitemOpen
  \bibfield  {author} {\bibinfo {author} {\bibfnamefont {L.~L.}\ \bibnamefont
  {Jensen}}, \bibinfo {author} {\bibfnamefont {J.~T.}\ \bibnamefont
  {Muckerman}}, \ and\ \bibinfo {author} {\bibfnamefont {M.~D.}\ \bibnamefont
  {Newton}},\ }\href {\doibase 10.1021/jp073554y} {\bibfield  {journal}
  {\bibinfo  {journal} {J. Phys. Chem. C}\ }\textbf {\bibinfo {volume} {112}},\
  \bibinfo {pages} {3439} (\bibinfo {year} {2008})}\BibitemShut {NoStop}%
\bibitem [{\citenamefont {Di~Valentin}(2010)}]{8Ub}%
  \BibitemOpen
  \bibfield  {author} {\bibinfo {author} {\bibfnamefont {C.}~\bibnamefont
  {Di~Valentin}},\ }\href {\doibase 10.1021/jp9112552} {\bibfield  {journal}
  {\bibinfo  {journal} {J. Phys. Chem. C}\ }\textbf {\bibinfo {volume} {114}},\
  \bibinfo {pages} {7054} (\bibinfo {year} {2010})}\BibitemShut {NoStop}%
\bibitem [{\citenamefont {Huda}\ \emph {et~al.}(2008)\citenamefont {Huda},
  \citenamefont {Yan}, \citenamefont {Wei},\ and\ \citenamefont
  {Al-Jassim}}]{Asymmetric}%
  \BibitemOpen
  \bibfield  {author} {\bibinfo {author} {\bibfnamefont {M.~N.}\ \bibnamefont
  {Huda}}, \bibinfo {author} {\bibfnamefont {Y.}~\bibnamefont {Yan}}, \bibinfo
  {author} {\bibfnamefont {S.-H.}\ \bibnamefont {Wei}}, \ and\ \bibinfo
  {author} {\bibfnamefont {M.~M.}\ \bibnamefont {Al-Jassim}},\ }\href {\doibase
  10.1103/PhysRevB.78.195204} {\bibfield  {journal} {\bibinfo  {journal} {Phys.
  Rev. B}\ }\textbf {\bibinfo {volume} {78}},\ \bibinfo {pages} {195204}
  (\bibinfo {year} {2008})}\BibitemShut {NoStop}%
\bibitem [{\citenamefont {Wei}\ \emph {et~al.}(1990)\citenamefont {Wei},
  \citenamefont {Ferreira}, \citenamefont {Bernard},\ and\ \citenamefont
  {Zunger}}]{SQS_PRB}%
  \BibitemOpen
  \bibfield  {author} {\bibinfo {author} {\bibfnamefont {S.-H.}\ \bibnamefont
  {Wei}}, \bibinfo {author} {\bibfnamefont {L.~G.}\ \bibnamefont {Ferreira}},
  \bibinfo {author} {\bibfnamefont {J.~E.}\ \bibnamefont {Bernard}}, \ and\
  \bibinfo {author} {\bibfnamefont {A.}~\bibnamefont {Zunger}},\ }\href
  {\doibase 10.1103/PhysRevB.42.9622} {\bibfield  {journal} {\bibinfo
  {journal} {Phys. Rev. B}\ }\textbf {\bibinfo {volume} {42}},\ \bibinfo
  {pages} {9622} (\bibinfo {year} {1990})}\BibitemShut {NoStop}%
\bibitem [{\citenamefont {Zunger}\ \emph {et~al.}(1990)\citenamefont {Zunger},
  \citenamefont {Wei}, \citenamefont {Ferreira},\ and\ \citenamefont
  {Bernard}}]{SQS_PRL}%
  \BibitemOpen
  \bibfield  {author} {\bibinfo {author} {\bibfnamefont {A.}~\bibnamefont
  {Zunger}}, \bibinfo {author} {\bibfnamefont {S.-H.}\ \bibnamefont {Wei}},
  \bibinfo {author} {\bibfnamefont {L.~G.}\ \bibnamefont {Ferreira}}, \ and\
  \bibinfo {author} {\bibfnamefont {J.~E.}\ \bibnamefont {Bernard}},\ }\href
  {\doibase 10.1103/PhysRevLett.65.353} {\bibfield  {journal} {\bibinfo
  {journal} {Phys. Rev. Lett.}\ }\textbf {\bibinfo {volume} {65}},\ \bibinfo
  {pages} {353} (\bibinfo {year} {1990})}\BibitemShut {NoStop}%
\bibitem [{\citenamefont {Ferhat}\ and\ \citenamefont
  {Bechstedt}(2002)}]{Bechstedt2002}%
  \BibitemOpen
  \bibfield  {author} {\bibinfo {author} {\bibfnamefont {M.}~\bibnamefont
  {Ferhat}}\ and\ \bibinfo {author} {\bibfnamefont {F.}~\bibnamefont
  {Bechstedt}},\ }\href {\doibase 10.1103/PhysRevB.65.075213} {\bibfield
  {journal} {\bibinfo  {journal} {Phys. Rev. B}\ }\textbf {\bibinfo {volume}
  {65}},\ \bibinfo {pages} {075213} (\bibinfo {year} {2002})}\BibitemShut
  {NoStop}%
\bibitem [{\citenamefont {\L{}opuszy\ifmmode~\acute{n}\else \'{n}\fi{}ski}\
  and\ \citenamefont {Majewski}(2012)}]{isovalent}%
  \BibitemOpen
  \bibfield  {author} {\bibinfo {author} {\bibfnamefont {M.}~\bibnamefont
  {\L{}opuszy\ifmmode~\acute{n}\else \'{n}\fi{}ski}}\ and\ \bibinfo {author}
  {\bibfnamefont {J.~A.}\ \bibnamefont {Majewski}},\ }\href {\doibase
  10.1103/PhysRevB.85.035211} {\bibfield  {journal} {\bibinfo  {journal} {Phys.
  Rev. B}\ }\textbf {\bibinfo {volume} {85}},\ \bibinfo {pages} {035211}
  (\bibinfo {year} {2012})}\BibitemShut {NoStop}%
\bibitem [{\citenamefont {M\"ader}\ and\ \citenamefont
  {Zunger}(1994)}]{Zunger1994}%
  \BibitemOpen
  \bibfield  {author} {\bibinfo {author} {\bibfnamefont {K.~A.}\ \bibnamefont
  {M\"ader}}\ and\ \bibinfo {author} {\bibfnamefont {A.}~\bibnamefont
  {Zunger}},\ }\href {\doibase http://dx.doi.org/10.1063/1.111403} {\bibfield
  {journal} {\bibinfo  {journal} {Applied Physics Letters}\ }\textbf {\bibinfo
  {volume} {64}},\ \bibinfo {pages} {2882} (\bibinfo {year}
  {1994})}\BibitemShut {NoStop}%
\bibitem [{\citenamefont {M\"ader}\ and\ \citenamefont
  {Zunger}(1995)}]{Zunger1995}%
  \BibitemOpen
  \bibfield  {author} {\bibinfo {author} {\bibfnamefont {K.~A.}\ \bibnamefont
  {M\"ader}}\ and\ \bibinfo {author} {\bibfnamefont {A.}~\bibnamefont
  {Zunger}},\ }\href {\doibase 10.1103/PhysRevB.51.10462} {\bibfield  {journal}
  {\bibinfo  {journal} {Phys. Rev. B}\ }\textbf {\bibinfo {volume} {51}},\
  \bibinfo {pages} {10462} (\bibinfo {year} {1995})}\BibitemShut {NoStop}%
\bibitem [{\citenamefont {de~Carvalho}\ \emph {et~al.}(2012)\citenamefont
  {de~Carvalho}, \citenamefont {Schleife}, \citenamefont {Furthm\"uller},\ and\
  \citenamefont {Bechstedt}}]{Bechstedt2012}%
  \BibitemOpen
  \bibfield  {author} {\bibinfo {author} {\bibfnamefont {L.~C.}\ \bibnamefont
  {de~Carvalho}}, \bibinfo {author} {\bibfnamefont {A.}~\bibnamefont
  {Schleife}}, \bibinfo {author} {\bibfnamefont {J.}~\bibnamefont
  {Furthm\"uller}}, \ and\ \bibinfo {author} {\bibfnamefont {F.}~\bibnamefont
  {Bechstedt}},\ }\href {\doibase 10.1103/PhysRevB.85.115121} {\bibfield
  {journal} {\bibinfo  {journal} {Phys. Rev. B}\ }\textbf {\bibinfo {volume}
  {85}},\ \bibinfo {pages} {115121} (\bibinfo {year} {2012})}\BibitemShut
  {NoStop}%
\bibitem [{\citenamefont {Saitta}\ \emph {et~al.}(1998)\citenamefont {Saitta},
  \citenamefont {de~Gironcoli},\ and\ \citenamefont {Baroni}}]{SQoSBaroni}%
  \BibitemOpen
  \bibfield  {author} {\bibinfo {author} {\bibfnamefont {A.~M.}\ \bibnamefont
  {Saitta}}, \bibinfo {author} {\bibfnamefont {S.}~\bibnamefont
  {de~Gironcoli}}, \ and\ \bibinfo {author} {\bibfnamefont {S.}~\bibnamefont
  {Baroni}},\ }\href {\doibase 10.1103/PhysRevLett.80.4939} {\bibfield
  {journal} {\bibinfo  {journal} {Phys. Rev. Lett.}\ }\textbf {\bibinfo
  {volume} {80}},\ \bibinfo {pages} {4939} (\bibinfo {year}
  {1998})}\BibitemShut {NoStop}%
\bibitem [{\citenamefont {Kim}\ and\ \citenamefont
  {Zunger}(2001)}]{SQoSZunger}%
  \BibitemOpen
  \bibfield  {author} {\bibinfo {author} {\bibfnamefont {K.}~\bibnamefont
  {Kim}}\ and\ \bibinfo {author} {\bibfnamefont {A.}~\bibnamefont {Zunger}},\
  }\href {\doibase 10.1103/PhysRevLett.86.2609} {\bibfield  {journal} {\bibinfo
   {journal} {Phys. Rev. Lett.}\ }\textbf {\bibinfo {volume} {86}},\ \bibinfo
  {pages} {2609} (\bibinfo {year} {2001})}\BibitemShut {NoStop}%
\bibitem [{\citenamefont {Li}\ \emph {et~al.}(2011)\citenamefont {Li},
  \citenamefont {Muckerman}, \citenamefont {Hybertsen},\ and\ \citenamefont
  {Allen}}]{LL}%
  \BibitemOpen
  \bibfield  {author} {\bibinfo {author} {\bibfnamefont {L.}~\bibnamefont
  {Li}}, \bibinfo {author} {\bibfnamefont {J.~T.}\ \bibnamefont {Muckerman}},
  \bibinfo {author} {\bibfnamefont {M.~S.}\ \bibnamefont {Hybertsen}}, \ and\
  \bibinfo {author} {\bibfnamefont {P.~B.}\ \bibnamefont {Allen}},\ }\href
  {\doibase 10.1103/PhysRevB.83.134202} {\bibfield  {journal} {\bibinfo
  {journal} {Phys. Rev. B}\ }\textbf {\bibinfo {volume} {83}},\ \bibinfo
  {pages} {134202} (\bibinfo {year} {2011})}\BibitemShut {NoStop}%
\bibitem [{\citenamefont {Sanchez}\ \emph {et~al.}(1984)\citenamefont
  {Sanchez}, \citenamefont {Ducastelle},\ and\ \citenamefont {Gratias}}]{CE1}%
  \BibitemOpen
  \bibfield  {author} {\bibinfo {author} {\bibfnamefont {J.}~\bibnamefont
  {Sanchez}}, \bibinfo {author} {\bibfnamefont {F.}~\bibnamefont {Ducastelle}},
  \ and\ \bibinfo {author} {\bibfnamefont {D.}~\bibnamefont {Gratias}},\ }\href
  {\doibase http://dx.doi.org/10.1016/0378-4371(84)90096-7} {\bibfield
  {journal} {\bibinfo  {journal} {Physica A}\ }\textbf {\bibinfo {volume}
  {128}},\ \bibinfo {pages} {334 } (\bibinfo {year} {1984})}\BibitemShut
  {NoStop}%
\bibitem [{\citenamefont {Connolly}\ and\ \citenamefont
  {Williams}(1983)}]{CE2}%
  \BibitemOpen
  \bibfield  {author} {\bibinfo {author} {\bibfnamefont {J.~W.~D.}\
  \bibnamefont {Connolly}}\ and\ \bibinfo {author} {\bibfnamefont {A.~R.}\
  \bibnamefont {Williams}},\ }\href {\doibase 10.1103/PhysRevB.27.5169}
  {\bibfield  {journal} {\bibinfo  {journal} {Phys. Rev. B}\ }\textbf {\bibinfo
  {volume} {27}},\ \bibinfo {pages} {5169} (\bibinfo {year}
  {1983})}\BibitemShut {NoStop}%
\bibitem [{\citenamefont {Fontaine}(1994)}]{CE3}%
  \BibitemOpen
  \bibfield  {author} {\bibinfo {author} {\bibfnamefont {D.~D.}\ \bibnamefont
  {Fontaine}},\ }in\ \href {\doibase
  http://dx.doi.org/10.1016/S0081-1947(08)60639-6} {\emph {\bibinfo {booktitle}
  {Solid State Physics}}},\ Vol.~\bibinfo {volume} {47},\ \bibinfo {editor}
  {edited by\ \bibinfo {editor} {\bibfnamefont {H.}~\bibnamefont {Ehrenreich}}\
  and\ \bibinfo {editor} {\bibfnamefont {D.}~\bibnamefont {Turnbull}}}\
  (\bibinfo  {publisher} {Academic Press},\ \bibinfo {year} {1994})\ pp.\
  \bibinfo {pages} {33 -- 176}\BibitemShut {NoStop}%
\bibitem [{\citenamefont {Ruban}\ and\ \citenamefont
  {Abrikosov}(2008)}]{CEECI}%
  \BibitemOpen
  \bibfield  {author} {\bibinfo {author} {\bibfnamefont {A.~V.}\ \bibnamefont
  {Ruban}}\ and\ \bibinfo {author} {\bibfnamefont {I.~A.}\ \bibnamefont
  {Abrikosov}},\ }\href {http://stacks.iop.org/0034-4885/71/i=4/a=046501}
  {\bibfield  {journal} {\bibinfo  {journal} {Reports on Progress in Physics}\
  }\textbf {\bibinfo {volume} {71}},\ \bibinfo {pages} {046501} (\bibinfo
  {year} {2008})}\BibitemShut {NoStop}%
\bibitem [{\citenamefont {Liu}\ \emph {et~al.}(2014)\citenamefont {Liu},
  \citenamefont {Pedroza}, \citenamefont {Misch}, \citenamefont
  {Fern\'{a}ndez-Serra},\ and\ \citenamefont {Allen}}]{Jian}%
  \BibitemOpen
  \bibfield  {author} {\bibinfo {author} {\bibfnamefont {J.}~\bibnamefont
  {Liu}}, \bibinfo {author} {\bibfnamefont {L.~S.}\ \bibnamefont {Pedroza}},
  \bibinfo {author} {\bibfnamefont {C.}~\bibnamefont {Misch}}, \bibinfo
  {author} {\bibfnamefont {M.~V.}\ \bibnamefont {Fern\'{a}ndez-Serra}}, \ and\
  \bibinfo {author} {\bibfnamefont {P.~B.}\ \bibnamefont {Allen}},\ }\href
  {http://stacks.iop.org/0953-8984/26/i=27/a=274204} {\bibfield  {journal}
  {\bibinfo  {journal} {J. Phys.: Condens. Matter}\ }\textbf {\bibinfo {volume}
  {26}},\ \bibinfo {pages} {274204} (\bibinfo {year} {2014})}\BibitemShut
  {NoStop}%
\bibitem [{\citenamefont {Liu}()}]{paper2}%
  \BibitemOpen
  \bibfield  {author} {\bibinfo {author} {\bibfnamefont {J.}~\bibnamefont
  {Liu}},\ }\href@noop {} {}\bibinfo {note} {(unpublished)}\BibitemShut
  {NoStop}%
\bibitem [{\citenamefont {van~de Walle}\ and\ \citenamefont
  {Asta}(2002)}]{ATAT1}%
  \BibitemOpen
  \bibfield  {author} {\bibinfo {author} {\bibfnamefont {A.}~\bibnamefont
  {van~de Walle}}\ and\ \bibinfo {author} {\bibfnamefont {M.}~\bibnamefont
  {Asta}},\ }\href {http://stacks.iop.org/0965-0393/10/i=5/a=304} {\bibfield
  {journal} {\bibinfo  {journal} {Modell. Simul. Mater. Sci. Eng.}\ }\textbf
  {\bibinfo {volume} {10}},\ \bibinfo {pages} {521} (\bibinfo {year}
  {2002})}\BibitemShut {NoStop}%
\bibitem [{\citenamefont {van~de Walle}\ \emph {et~al.}(2002)\citenamefont
  {van~de Walle}, \citenamefont {Asta},\ and\ \citenamefont {Ceder}}]{ATAT2}%
  \BibitemOpen
  \bibfield  {author} {\bibinfo {author} {\bibfnamefont {A.}~\bibnamefont
  {van~de Walle}}, \bibinfo {author} {\bibfnamefont {M.}~\bibnamefont {Asta}},
  \ and\ \bibinfo {author} {\bibfnamefont {G.}~\bibnamefont {Ceder}},\ }\href
  {\doibase http://dx.doi.org/10.1016/S0364-5916(02)80006-2} {\bibfield
  {journal} {\bibinfo  {journal} {Calphad}\ }\textbf {\bibinfo {volume} {26}},\
  \bibinfo {pages} {539 } (\bibinfo {year} {2002})}\BibitemShut {NoStop}%
\bibitem [{\citenamefont {van~de Walle}\ and\ \citenamefont
  {Ceder}(2002{\natexlab{a}})}]{ATAT3}%
  \BibitemOpen
  \bibfield  {author} {\bibinfo {author} {\bibfnamefont {A.}~\bibnamefont
  {van~de Walle}}\ and\ \bibinfo {author} {\bibfnamefont {G.}~\bibnamefont
  {Ceder}},\ }\href {\doibase 10.1361/105497102770331596} {\bibfield  {journal}
  {\bibinfo  {journal} {J. Phase Equilib.}\ }\textbf {\bibinfo {volume} {23}},\
  \bibinfo {pages} {348} (\bibinfo {year} {2002}{\natexlab{a}})}\BibitemShut
  {NoStop}%
\bibitem [{SM()}]{SM}%
  \BibitemOpen
  \href@noop {} {}\bibinfo {note} {See Supplemental Material for the special
  structures.}\BibitemShut {Stop}%
\bibitem [{\citenamefont {Giannozzi}\ \emph {et~al.}(2009)\citenamefont
  {Giannozzi}, \citenamefont {Baroni}, \citenamefont {Bonini}, \citenamefont
  {Calandra}, \citenamefont {Car}, \citenamefont {Cavazzoni}, \citenamefont
  {Ceresoli}, \citenamefont {Chiarotti}, \citenamefont {Cococcioni},
  \citenamefont {Dabo}, \citenamefont {Corso}, \citenamefont {de~Gironcoli},
  \citenamefont {Fabris}, \citenamefont {Fratesi}, \citenamefont {Gebauer},
  \citenamefont {Gerstmann}, \citenamefont {Gougoussis}, \citenamefont
  {Kokalj}, \citenamefont {Lazzeri}, \citenamefont {Martin-Samos},
  \citenamefont {Marzari}, \citenamefont {Mauri}, \citenamefont {Mazzarello},
  \citenamefont {Paolini}, \citenamefont {Pasquarello}, \citenamefont
  {Paulatto}, \citenamefont {Sbraccia}, \citenamefont {Scandolo}, \citenamefont
  {Sclauzero}, \citenamefont {Seitsonen}, \citenamefont {Smogunov},
  \citenamefont {Umari},\ and\ \citenamefont {Wentzcovitch}}]{QE}%
  \BibitemOpen
  \bibfield  {author} {\bibinfo {author} {\bibfnamefont {P.}~\bibnamefont
  {Giannozzi}}, \bibinfo {author} {\bibfnamefont {S.}~\bibnamefont {Baroni}},
  \bibinfo {author} {\bibfnamefont {N.}~\bibnamefont {Bonini}}, \bibinfo
  {author} {\bibfnamefont {M.}~\bibnamefont {Calandra}}, \bibinfo {author}
  {\bibfnamefont {R.}~\bibnamefont {Car}}, \bibinfo {author} {\bibfnamefont
  {C.}~\bibnamefont {Cavazzoni}}, \bibinfo {author} {\bibfnamefont
  {D.}~\bibnamefont {Ceresoli}}, \bibinfo {author} {\bibfnamefont {G.~L.}\
  \bibnamefont {Chiarotti}}, \bibinfo {author} {\bibfnamefont {M.}~\bibnamefont
  {Cococcioni}}, \bibinfo {author} {\bibfnamefont {I.}~\bibnamefont {Dabo}},
  \bibinfo {author} {\bibfnamefont {A.~D.}\ \bibnamefont {Corso}}, \bibinfo
  {author} {\bibfnamefont {S.}~\bibnamefont {de~Gironcoli}}, \bibinfo {author}
  {\bibfnamefont {S.}~\bibnamefont {Fabris}}, \bibinfo {author} {\bibfnamefont
  {G.}~\bibnamefont {Fratesi}}, \bibinfo {author} {\bibfnamefont
  {R.}~\bibnamefont {Gebauer}}, \bibinfo {author} {\bibfnamefont
  {U.}~\bibnamefont {Gerstmann}}, \bibinfo {author} {\bibfnamefont
  {C.}~\bibnamefont {Gougoussis}}, \bibinfo {author} {\bibfnamefont
  {A.}~\bibnamefont {Kokalj}}, \bibinfo {author} {\bibfnamefont
  {M.}~\bibnamefont {Lazzeri}}, \bibinfo {author} {\bibfnamefont
  {L.}~\bibnamefont {Martin-Samos}}, \bibinfo {author} {\bibfnamefont
  {N.}~\bibnamefont {Marzari}}, \bibinfo {author} {\bibfnamefont
  {F.}~\bibnamefont {Mauri}}, \bibinfo {author} {\bibfnamefont
  {R.}~\bibnamefont {Mazzarello}}, \bibinfo {author} {\bibfnamefont
  {S.}~\bibnamefont {Paolini}}, \bibinfo {author} {\bibfnamefont
  {A.}~\bibnamefont {Pasquarello}}, \bibinfo {author} {\bibfnamefont
  {L.}~\bibnamefont {Paulatto}}, \bibinfo {author} {\bibfnamefont
  {C.}~\bibnamefont {Sbraccia}}, \bibinfo {author} {\bibfnamefont
  {S.}~\bibnamefont {Scandolo}}, \bibinfo {author} {\bibfnamefont
  {G.}~\bibnamefont {Sclauzero}}, \bibinfo {author} {\bibfnamefont {A.~P.}\
  \bibnamefont {Seitsonen}}, \bibinfo {author} {\bibfnamefont {A.}~\bibnamefont
  {Smogunov}}, \bibinfo {author} {\bibfnamefont {P.}~\bibnamefont {Umari}}, \
  and\ \bibinfo {author} {\bibfnamefont {R.~M.}\ \bibnamefont {Wentzcovitch}},\
  }\href {http://stacks.iop.org/0953-8984/21/i=39/a=395502} {\bibfield
  {journal} {\bibinfo  {journal} {J. Phys.: Condens. Matter}\ }\textbf
  {\bibinfo {volume} {21}},\ \bibinfo {pages} {395502} (\bibinfo {year}
  {2009})}\BibitemShut {NoStop}%
\bibitem [{\citenamefont {Perdew}\ \emph {et~al.}(2008)\citenamefont {Perdew},
  \citenamefont {Ruzsinszky}, \citenamefont {Csonka}, \citenamefont {Vydrov},
  \citenamefont {Scuseria}, \citenamefont {Constantin}, \citenamefont {Zhou},\
  and\ \citenamefont {Burke}}]{PBEsol}%
  \BibitemOpen
  \bibfield  {author} {\bibinfo {author} {\bibfnamefont {J.~P.}\ \bibnamefont
  {Perdew}}, \bibinfo {author} {\bibfnamefont {A.}~\bibnamefont {Ruzsinszky}},
  \bibinfo {author} {\bibfnamefont {G.~I.}\ \bibnamefont {Csonka}}, \bibinfo
  {author} {\bibfnamefont {O.~A.}\ \bibnamefont {Vydrov}}, \bibinfo {author}
  {\bibfnamefont {G.~E.}\ \bibnamefont {Scuseria}}, \bibinfo {author}
  {\bibfnamefont {L.~A.}\ \bibnamefont {Constantin}}, \bibinfo {author}
  {\bibfnamefont {X.}~\bibnamefont {Zhou}}, \ and\ \bibinfo {author}
  {\bibfnamefont {K.}~\bibnamefont {Burke}},\ }\href {\doibase
  10.1103/PhysRevLett.100.136406} {\bibfield  {journal} {\bibinfo  {journal}
  {Phys. Rev. Lett.}\ }\textbf {\bibinfo {volume} {100}},\ \bibinfo {pages}
  {136406} (\bibinfo {year} {2008})}\BibitemShut {NoStop}%
\bibitem [{\citenamefont {Bl\"ochl}(1994)}]{PAW}%
  \BibitemOpen
  \bibfield  {author} {\bibinfo {author} {\bibfnamefont {P.~E.}\ \bibnamefont
  {Bl\"ochl}},\ }\href {\doibase 10.1103/PhysRevB.50.17953} {\bibfield
  {journal} {\bibinfo  {journal} {Phys. Rev. B}\ }\textbf {\bibinfo {volume}
  {50}},\ \bibinfo {pages} {17953} (\bibinfo {year} {1994})}\BibitemShut
  {NoStop}%
\bibitem [{\citenamefont {Corso}(2014)}]{PSL}%
  \BibitemOpen
  \bibfield  {author} {\bibinfo {author} {\bibfnamefont {A.~D.}\ \bibnamefont
  {Corso}},\ }\href {\doibase
  http://dx.doi.org/10.1016/j.commatsci.2014.07.043} {\bibfield  {journal}
  {\bibinfo  {journal} {Computational Materials Science}\ }\textbf {\bibinfo
  {volume} {95}},\ \bibinfo {pages} {337 } (\bibinfo {year}
  {2014})}\BibitemShut {NoStop}%
\bibitem [{\citenamefont {Denton}\ and\ \citenamefont
  {Ashcroft}(1991)}]{Vegard}%
  \BibitemOpen
  \bibfield  {author} {\bibinfo {author} {\bibfnamefont {A.~R.}\ \bibnamefont
  {Denton}}\ and\ \bibinfo {author} {\bibfnamefont {N.~W.}\ \bibnamefont
  {Ashcroft}},\ }\href {\doibase 10.1103/PhysRevA.43.3161} {\bibfield
  {journal} {\bibinfo  {journal} {Phys. Rev. A}\ }\textbf {\bibinfo {volume}
  {43}},\ \bibinfo {pages} {3161} (\bibinfo {year} {1991})}\BibitemShut
  {NoStop}%
\bibitem [{\citenamefont {Alf\`{e}}(2009)}]{PHON}%
  \BibitemOpen
  \bibfield  {author} {\bibinfo {author} {\bibfnamefont {D.}~\bibnamefont
  {Alf\`{e}}},\ }\href {\doibase http://dx.doi.org/10.1016/j.cpc.2009.03.010}
  {\bibfield  {journal} {\bibinfo  {journal} {Comput. Phys. Commun.}\ }\textbf
  {\bibinfo {volume} {180}},\ \bibinfo {pages} {2622 } (\bibinfo {year}
  {2009})}\BibitemShut {NoStop}%
\bibitem [{\citenamefont {Soler}\ \emph {et~al.}(2002)\citenamefont {Soler},
  \citenamefont {Artacho}, \citenamefont {Gale}, \citenamefont {Garc\'{i}a},
  \citenamefont {Junquera}, \citenamefont {Ordej\'{o}n},\ and\ \citenamefont
  {S\'{a}nchez-Portal}}]{SIESTA}%
  \BibitemOpen
  \bibfield  {author} {\bibinfo {author} {\bibfnamefont {J.~M.}\ \bibnamefont
  {Soler}}, \bibinfo {author} {\bibfnamefont {E.}~\bibnamefont {Artacho}},
  \bibinfo {author} {\bibfnamefont {J.~D.}\ \bibnamefont {Gale}}, \bibinfo
  {author} {\bibfnamefont {A.}~\bibnamefont {Garc\'{i}a}}, \bibinfo {author}
  {\bibfnamefont {J.}~\bibnamefont {Junquera}}, \bibinfo {author}
  {\bibfnamefont {P.}~\bibnamefont {Ordej\'{o}n}}, \ and\ \bibinfo {author}
  {\bibfnamefont {D.}~\bibnamefont {S\'{a}nchez-Portal}},\ }\href
  {http://stacks.iop.org/0953-8984/14/i=11/a=302} {\bibfield  {journal}
  {\bibinfo  {journal} {J. Phys.: Condens. Matter}\ }\textbf {\bibinfo {volume}
  {14}},\ \bibinfo {pages} {2745} (\bibinfo {year} {2002})}\BibitemShut
  {NoStop}%
\bibitem [{hom()}]{homepage}%
  \BibitemOpen
  \href@noop {} {\ }\BibitemShut {NoStop}%
\bibitem [{\citenamefont {Agrawal}\ and\ \citenamefont
  {Espinosa}(2011)}]{Ga_pseudo}%
  \BibitemOpen
  \bibfield  {author} {\bibinfo {author} {\bibfnamefont {R.}~\bibnamefont
  {Agrawal}}\ and\ \bibinfo {author} {\bibfnamefont {H.~D.}\ \bibnamefont
  {Espinosa}},\ }\href {\doibase 10.1021/nl104004d} {\bibfield  {journal}
  {\bibinfo  {journal} {Nano Lett.}\ }\textbf {\bibinfo {volume} {11}},\
  \bibinfo {pages} {786} (\bibinfo {year} {2011})}\BibitemShut {NoStop}%
\bibitem [{\citenamefont {Dudarev}\ \emph {et~al.}(1998)\citenamefont
  {Dudarev}, \citenamefont {Botton}, \citenamefont {Savrasov}, \citenamefont
  {Humphreys},\ and\ \citenamefont {Sutton}}]{Dudarev}%
  \BibitemOpen
  \bibfield  {author} {\bibinfo {author} {\bibfnamefont {S.~L.}\ \bibnamefont
  {Dudarev}}, \bibinfo {author} {\bibfnamefont {G.~A.}\ \bibnamefont {Botton}},
  \bibinfo {author} {\bibfnamefont {S.~Y.}\ \bibnamefont {Savrasov}}, \bibinfo
  {author} {\bibfnamefont {C.~J.}\ \bibnamefont {Humphreys}}, \ and\ \bibinfo
  {author} {\bibfnamefont {A.~P.}\ \bibnamefont {Sutton}},\ }\href {\doibase
  10.1103/PhysRevB.57.1505} {\bibfield  {journal} {\bibinfo  {journal} {Phys.
  Rev. B}\ }\textbf {\bibinfo {volume} {57}},\ \bibinfo {pages} {1505}
  (\bibinfo {year} {1998})}\BibitemShut {NoStop}%
\bibitem [{\citenamefont {Janotti}\ \emph {et~al.}(2006)\citenamefont
  {Janotti}, \citenamefont {Segev},\ and\ \citenamefont {Van~de
  Walle}}]{plusU}%
  \BibitemOpen
  \bibfield  {author} {\bibinfo {author} {\bibfnamefont {A.}~\bibnamefont
  {Janotti}}, \bibinfo {author} {\bibfnamefont {D.}~\bibnamefont {Segev}}, \
  and\ \bibinfo {author} {\bibfnamefont {C.~G.}\ \bibnamefont {Van~de Walle}},\
  }\href {\doibase 10.1103/PhysRevB.74.045202} {\bibfield  {journal} {\bibinfo
  {journal} {Phys. Rev. B}\ }\textbf {\bibinfo {volume} {74}},\ \bibinfo
  {pages} {045202} (\bibinfo {year} {2006})}\BibitemShut {NoStop}%
\bibitem [{\citenamefont {Baroni}\ \emph {et~al.}(2001)\citenamefont {Baroni},
  \citenamefont {de~Gironcoli}, \citenamefont {Dal~Corso},\ and\ \citenamefont
  {Giannozzi}}]{linear_response}%
  \BibitemOpen
  \bibfield  {author} {\bibinfo {author} {\bibfnamefont {S.}~\bibnamefont
  {Baroni}}, \bibinfo {author} {\bibfnamefont {S.}~\bibnamefont
  {de~Gironcoli}}, \bibinfo {author} {\bibfnamefont {A.}~\bibnamefont
  {Dal~Corso}}, \ and\ \bibinfo {author} {\bibfnamefont {P.}~\bibnamefont
  {Giannozzi}},\ }\href {\doibase 10.1103/RevModPhys.73.515} {\bibfield
  {journal} {\bibinfo  {journal} {Rev. Mod. Phys.}\ }\textbf {\bibinfo {volume}
  {73}},\ \bibinfo {pages} {515} (\bibinfo {year} {2001})}\BibitemShut
  {NoStop}%
\bibitem [{\citenamefont {Marques}\ \emph {et~al.}(2011)\citenamefont
  {Marques}, \citenamefont {Vidal}, \citenamefont {Oliveira}, \citenamefont
  {Reining},\ and\ \citenamefont {Botti}}]{alpha}%
  \BibitemOpen
  \bibfield  {author} {\bibinfo {author} {\bibfnamefont {M.~A.~L.}\
  \bibnamefont {Marques}}, \bibinfo {author} {\bibfnamefont {J.}~\bibnamefont
  {Vidal}}, \bibinfo {author} {\bibfnamefont {M.~J.~T.}\ \bibnamefont
  {Oliveira}}, \bibinfo {author} {\bibfnamefont {L.}~\bibnamefont {Reining}}, \
  and\ \bibinfo {author} {\bibfnamefont {S.}~\bibnamefont {Botti}},\ }\href
  {\doibase 10.1103/PhysRevB.83.035119} {\bibfield  {journal} {\bibinfo
  {journal} {Phys. Rev. B}\ }\textbf {\bibinfo {volume} {83}},\ \bibinfo
  {pages} {035119} (\bibinfo {year} {2011})}\BibitemShut {NoStop}%
\bibitem [{\citenamefont {Hanada}(2009)}]{property}%
  \BibitemOpen
  \bibfield  {author} {\bibinfo {author} {\bibfnamefont {T.}~\bibnamefont
  {Hanada}},\ }in\ \href {\doibase 10.1007/978-3-540-88847-5_1} {\emph
  {\bibinfo {booktitle} {Oxide and Nitride Semiconductors}}},\ \bibinfo
  {series} {Advances in Materials Research}, Vol.~\bibinfo {volume} {12},\
  \bibinfo {editor} {edited by\ \bibinfo {editor} {\bibfnamefont
  {T.}~\bibnamefont {Yao}}\ and\ \bibinfo {editor} {\bibfnamefont {S.-K.}\
  \bibnamefont {Hong}}}\ (\bibinfo  {publisher} {Springer Berlin Heidelberg},\
  \bibinfo {year} {2009})\ pp.\ \bibinfo {pages} {1--19}\BibitemShut {NoStop}%
\bibitem [{\citenamefont {Perdew}\ \emph {et~al.}(1996)\citenamefont {Perdew},
  \citenamefont {Burke},\ and\ \citenamefont {Ernzerhof}}]{PBE}%
  \BibitemOpen
  \bibfield  {author} {\bibinfo {author} {\bibfnamefont {J.~P.}\ \bibnamefont
  {Perdew}}, \bibinfo {author} {\bibfnamefont {K.}~\bibnamefont {Burke}}, \
  and\ \bibinfo {author} {\bibfnamefont {M.}~\bibnamefont {Ernzerhof}},\ }\href
  {\doibase 10.1103/PhysRevLett.77.3865} {\bibfield  {journal} {\bibinfo
  {journal} {Phys. Rev. Lett.}\ }\textbf {\bibinfo {volume} {77}},\ \bibinfo
  {pages} {3865} (\bibinfo {year} {1996})}\BibitemShut {NoStop}%
\bibitem [{\citenamefont {Brown}(2009)}]{BVM}%
  \BibitemOpen
  \bibfield  {author} {\bibinfo {author} {\bibfnamefont {I.~D.}\ \bibnamefont
  {Brown}},\ }\href {\doibase 10.1021/cr900053k} {\bibfield  {journal}
  {\bibinfo  {journal} {Chem. Rev.}\ }\textbf {\bibinfo {volume} {109}},\
  \bibinfo {pages} {6858} (\bibinfo {year} {2009})}\BibitemShut {NoStop}%
\bibitem [{\citenamefont {van~de Walle}\ and\ \citenamefont
  {Ceder}(2002{\natexlab{b}})}]{RMP}%
  \BibitemOpen
  \bibfield  {author} {\bibinfo {author} {\bibfnamefont {A.}~\bibnamefont
  {van~de Walle}}\ and\ \bibinfo {author} {\bibfnamefont {G.}~\bibnamefont
  {Ceder}},\ }\href {\doibase 10.1103/RevModPhys.74.11} {\bibfield  {journal}
  {\bibinfo  {journal} {Rev. Mod. Phys.}\ }\textbf {\bibinfo {volume} {74}},\
  \bibinfo {pages} {11} (\bibinfo {year} {2002}{\natexlab{b}})}\BibitemShut
  {NoStop}%
\end{thebibliography}%

\end{document}